\documentstyle[12pt,aasms4]{article}

\newcommand\th{\thinspace}
\newcommand\kms{\ifmmode{\rm km\th s^{-1}}\else km\th s$^{-1}$\fi}
\newcommand\msun{\ifmmode{M_{\odot}}\else $M_{\odot}$\fi}
\newcommand\rsun{\ifmmode{R_{\odot}}\else $R_{\odot}$\fi}
\newcommand \p  {$\pm$}

\def\vrotsini{$v_{\rm rot}$ sin~$i$}

\begin{document}

\title{METAL-POOR FIELD BLUE STRAGGLERS: MORE EVIDENCE
FOR MASS TRANSFER\footnotemark}

\footnotetext{Some of the results presented here used observations
made with the Multiple Mirror Telescope, a joint facility of the
Smithsonian Institution and the University of Arizona}

\author{Bruce W. Carney}
\affil{Department of Physics \& Astronomy, University of North Carolina,
Chapel Hill, NC 27599-3255; bruce@physics.unc.edu}
\author{David W. Latham}
\affil{Harvard-Smithsonian Center for Astrophysics, 60 Garden Street,
Cambridge, MA 02138;  dlatham@cfa.harvard.edu}
\author{John B. Laird}
\affil{Department of Physics \& Astronomy, Bowling Green State University,
Bowling Green, OH 43403; laird@tycho.bgsu.edu}

\clearpage

\begin{abstract} 

We report radial velocity studies of five candidate metal-poor field
blue stragglers, all known to be deficient in lithium.
Four of the five stars are single-lined spectroscopic binaries, with
periods ranging from 302 to 840 days, and low orbital eccentricities,
in agreement with similar behavior found for other blue straggler
candidates by \markcite{PS00}Preston \& Sneden (2000) and 
\markcite{CLLGM}Carney et al.\ (2001).
The limited data available for lithium abundances indicate that all
blue straggler binaries have depleted lithium, but that constant
velocity stars generally have normal lithium abundances. This
suggests that the ``lithium gap" for hot metal-poor main sequence
stars may not exist or lies at higher temperatures than found
in the Hyades. Our results and those of \markcite{PS00}Preston \& Sneden (2000)
show higher values of \vrotsini\ for the binary stars
than those of comparable temperature constant velocity stars.
The orbital periods are too long for tidal effects to be important,
implying that spin-up during mass transfer when the
orbital separations and periods were smaller
is the cause of the enhanced rotation. The mass function
distribution is steeper for the blue straggler
binary stars than that of lower mass single-lined
spectroscopic binaries, indicating a narrower range in secondary
masses. We argue that if all secondaries are white dwarfs with
the same mass, it is probably around 0.55 \msun. The models
of \markcite{rappaport95}Rappaport et al.\ (1995), 
applied to white dwarf secondaries,
suggest that the orbital elements of all metal-poor 
binary blue stragglers are consistent
with stable mass transfer, with the possible exception of G202-65.

\end{abstract}


\section{INTRODUCTION}

The ``blue stragglers" were first identified by 
\markcite{S53}Sandage (1953) in the globular cluster M3. Since
their discovery, blue stragglers have been found in many
other globular clusters, and in open clusters as well. The frequency
of blue stragglers is also correlated with position in clusters,
being generally higher in the denser, more dynamically evolved portions,
indicating that stellar interactions, perhaps stellar mergers or hardening
of binaries leading to mass transfer, may be the cause or at least one
of the causes of the phenomenon (\markcite{mateo90}Mateo et al.\ 1990). 
\markcite{Hut92}Hut et al.\ (1992),
\markcite{S93}Stryker (1993), \markcite{L93}Livio
(1993), \markcite{T93}Trimble (1993), \markcite{B95}Bailyn (1995), and
\markcite{L96}Leonard (1996) have provided  extensive reviews of
blue stragglers in globular clusters.

Some globular clusters show declining frequencies (per unit
luminosity) of blue stragglers in their lower density outer
regions, but others (M3: Ferraro et al.\ 1993, 1997;
M55: Zaggia et al.\ 1997; 47~Tuc: Ferraro et al.\ 2004 and
Mapelli et al.\ 2004) show an enhanced frequency in the outermost,
lowest density regions. This is explained qualitatively
by invoking two different mechanisms for the creation
of blue stragglers in clusters. In the central regions,
dynamical effects involving collisions may enhance the probabilities for
creating blue stragglers (\markcite{SB99}Sills \& Bailyn 1999).
In the outer regions, normal stellar evolution may be
responsible for their creation.

A variety of stellar evolution explanations 
have been proposed to explain such unusual
systems, one of which (mass transfer: \markcite{M64}McCrea 1964) does
not require any alterations to standard stellar evolution theory.
But two explanations (pulsation-driven mass loss: 
\markcite{WBS87}Willson, Bowen, \& Struck-Marcell 1987; internal mixing:
\markcite{W79a,W79b}Wheeler 1979a,b;
\markcite{SW80}Saio \& Wheeler 1980) do require significant
modifications and make the understanding of the blue straggler
phenomenon of high importance. The mass transfer model
is perhaps favored by the increasing blue straggler frequencies
seen in the outer regions of the above clusters because
that is where binary systems are most likely to survive. 
\markcite{mapelli04}Mapelli et al.\ (2004), for example,
were able to reproduce the blue straggler binary frequency
distribution in 47~Tuc using a combination of collisional
creation of blue stragglers and survival of mass transfer
systems. The collisions, as expected, dominate the central
regions, while the outer regions' blue stragglers arise both
from collisions and mass transfer in binary systems.

\markcite{PS00}Preston \& Sneden (2000; hereafter PS2000)
expanded on the earlier work of \markcite{PBS94}Preston, Beers,
\& Shectman (1994; hereafter PBS) and argued that
{\em field} blue stragglers are created almost solely by
the mass transfer process. Not only does this make good
qualitative sense, but they supported the argument by noting
that field blue stragglers are much more common per unit
luminosity than are blue stragglers in the lower-density
regions of globular clusters. As PS2000 noted, this is
consistent with the idea that even in those low density
regions of clusters, binary systems may have been destroyed, 
thereby lowering the numbers
of blue stragglers. 

In clusters, we must keep in mind that even the stars
found in the low-density outer regions may be vulnerable
to stellar encounters if their intra-cluster motions carry
them into the central higher density regions.
Field blue stragglers therefore
provide the ``cleanest" sample for the study of the
mechanism that may create such stellar oddities 
via stellar evolution, independent
of environment.
The distributions of such observables as binary stars'
periods and orbital eccentricities are therefore
much less likely to have been among field stars than
in clusters.

As we describe
below, however, field star samples have introduced another puzzle, one
at least as intriguing as the blue straggler phenomenon itself. We
argued \markcite{CLLGM}(Carney et al.\ 2001; hereafter CLLGM) and argue again here 
that binary star evolution and mass transfer is one path for the creation
of blue stragglers, and apparently the most common among field stars. 
However, some metal-poor main sequence field stars that are hotter
than globular cluster main sequence turn-offs are not binaries,
but could be explained as 
the partial remnants of an accreted
dwarf satellite galaxy whose star formation continued over
a long period of time. Following PBS,
we distinguish between stars with delayed evolution
and those that are simply young stars. We regard the blue straggler
phenomenon as the process(es) that delay(s) normal stellar evolution.

\markcite{PBS94}PBS identified a significant number of 
field metal-poor stars bluer than
comparable metallicity globular cluster main sequence 
turn-off stars: candidate blue
stragglers, in other words. 
\markcite{PL98}Preston \& Landolt (1998) found that the star 
CS~$22966-043$ is a very metal-poor ([Fe/H] $\approx -2.4$) 
SX~Phe pulsating variable, and is also a binary star with
an orbital period of 431 days and a relatively low orbital
eccentricity of 0.10. (SX Phe variables lie in the instability strip's
extension to the main sequence domain, and thus represent the hotter
blue stragglers in globular clusters: see \markcite{NNLL94}Nemec, Nemec,
\& Lutz 1994
for a discussion and lists of such stars.) The orbital properties are
a vital clue to its origin because the normal orbital eccentricity of main sequence
binary stars is much higher, with $<$e$>$ close to 0.37 
(\markcite{DM91}Duquennoy \& Mayor 1991,
\markcite{XV}Latham et al.\ 2002), independent of metallicity. An extensive study
of blue metal-poor field stars by 
PS2000 found a surprisingly high fraction of binary stars, 
which led them to suggest that
many of them were field blue stragglers. Many of these stars showed
periods of hundreds of days with low orbital eccentricities, as in
the case of CS~$22966-043$. 
\markcite{CLLGM}CLLGM identified 
ten metal-poor field blue straggler candidate stars
from the large survey of \markcite{XII}Carney et al.\ (1994), based on
colors that were slightly to significantly
bluer than globular cluster main sequence turn-offs (at similar metallicities). They
found six of the stars to have periods in this same range, 167 to 844 days,
and low orbital eccentricities ($e \leq\ 0.26$; $<e> = 0.11$). They argued
that these periods and eccentricities are consistent with mass transfer.
They noted that 
mass transfer systems such as CH, subgiant CH, and dwarf carbon stars
likewise have very high binary fractions and that those binary systems
have similarly long orbital periods and low orbital eccentricities.
\markcite{SPC2003}Sneden, Preston, \& Cowan (2003; hereafter SPC2003) have strengthened
the argument for mass transfer in their recent abundance analyses of
six blue metal-poor stars from the \markcite{PS00}PS2000 program, finding that the
three constant velocity stars have normal element-to-iron abundance
ratios for metal-poor stars, but that three binary stars show 
major enhancements of carbon and $s$-process elements, indicative
of mass transfer from a highly-evolved AGB star.
In summary, then, mass transfer appears to be the primary cause
of the blue straggler phenomenon among field stars. Therefore, it may also be
a major contributor to the formation of blue stragglers in clusters,
although the situation there is more complicated by induced dynamical
evolution of binary stars as well as outright capture and merger of stars.

Since the orbital characteristics
are such an important component of the mass transfer model, 
more velocity data of known field blue stragglers and
identification of additional candidates and their binary frequency and orbital
characteristics are highly desirable. It is also important to employ
consistent techniques for identifying candidate blue stragglers and
for determining the orbital elements of the binary stars.
In the latter case, we note that our criteria for identification of binaries
and acceptance of final orbital solutions 
are more strict than those of some investigators (see 
\markcite{VI}Latham et al.\ 1988, \markcite{XI}Latham
et al.\ 1992, and \markcite{XVI}Latham et al.\ 2002). This
has compelled us to re-evaluate the orbital solutions available
in the literature, using the published velocity data, in order to
re-analyze all the velocity data in a consistent manner.
Thus, while \markcite{PS00}PS2000 found 42 binaries among their 62 program stars, 
we could unambiguously identify only
29 stars as being binaries using their published velocities
and our software and experience. The other 13 stars may indeed be binaries,
but our criteria and experience lead us to
believe that additional radial velocities are
needed to confirm such identifications.
We also note that for some stars, \markcite{PS00}PS2000 did not
have enough time coverage to obtain a unique orbital
period and offered alternative solutions when necessary. 
We have retained for use in this paper
only the ten stars from \markcite{PS00}PS2000 whose orbital elements satisfy our
criteria. In all these cases, our orbital solutions are indistinguishable
from those obtained by \markcite{PS00}PS2000. 
\markcite{SPC2003}SPC2003 have also addressed this 
point with additional
radial velocity data for selected stars. We are confident that additional
observations will resolve any remaining uncertainties. We add that
G.\ Preston (priv. comm.) has obtained additional velocities for
CS~$29497-030$ and CS~$29509-027$, and those results confirm
the orbital solutions for those stars presented by \markcite{SPC2003}SPC2003. 

For this paper, we have chosen to concentrate our efforts on seeking
additional stars, rather than refinements of existing orbital solutions presented
by \markcite{PS00}PS2000. The exciting results 
of \markcite{SPC2003}SPC2003 provides another motivation: a larger
sample, and of brighter stars, could enable further progress in the search
for chemical signatures of mass transfer, and, perhaps, an understanding of
different mass transfer mechanisms (e.g., red giant or AGB donor star?). Finally,
we have not lost sight of perhaps the most intriguing discovery of \markcite{PBS94}PBS:
they argued that many of their stars are merely masquerading as blue stragglers
and that they may, in fact, be bona fide young, metal-poor stars, accreted perhaps
from a small dwarf galaxy which had experienced a prolonged star formation history
or perhaps a relatively recent burst of star formation. Candidate field
blue straggler stars may therefore provide clues to both stellar evolution and
the evolution of our Galaxy.

\section{SELECTION OF FIELD BLUE STRAGGLERS \& SOME 
INTERESTING PROBLEMS THEREIN}

\subsection{Finding Candidate Blue Stragglers}

We have two challenges in identifying metal-poor blue stragglers in the
field population. The first is the rarity of such stars. Large samples
of field stars must be studied to find those few that are bluer than
the main sequence turn-off of comparable metallicity globular clusters,
and we must also be assured that they are main sequence stars and not
the more luminous blue horizontal branch stars.
Thus accurate colors, reddenings, 
gravities, and metallicities are required for
large numbers of stars. This is why the two major recent studies of
metal-poor field blue stragglers have emerged from the ``BMP" (Blue
Metal Poor) studies like that of \markcite{PBS94}PBS 
and \markcite{PS00}PS2000, and the ``Carney-Latham-Laird-Aguilar"
survey (\markcite{XII}Carney et al.\ 1994; \markcite{CLLGM}CLLGM).
Such methods pick up both true blue stragglers as well as any
truly young metal-poor stars, which may have been acquired by the
Galaxy during a minor merger event (\markcite{PBS94}PBS), as noted above. 

In their study of 62 blue metal-poor stars, many
of which qualify as blue straggler candidates
using our color criterion, \markcite{PS00}PS2000 found an
abnormally high fraction of binary stars, roughly four times
higher than those of the disk and halo 
populations (\markcite{DM91}Duquennoy \& Mayor 1991;
\markcite{VI,XVI}Latham et al.\ 1988, 2002). 
On the other hand, the frequency of
double-lined systems was unusually low, suggesting a population of
underluminous secondary stars, possibly white dwarfs. 
Such evidence, plus a low fraction of binary
systems with short orbital periods, and a very high fraction of
binary systems with low orbital eccentricities, suggested to \markcite{PS00}PS2000
that their large sample of blue metal-poor stars is a mixture
of both relatively young stars, with, presumably, a normal frequency
of binary stars and with orbital elements like those of Galactic
disk and halo stars, plus a significant number of blue stragglers
whose origin is most readily explained by mass transfer.

\subsection{The Role of Lithium}

Lithium is a very ``fragile'' element, vulnerable to proton
capture at the relatively low temperature of
order $2.5 \times 10^6$~K.  Cool stars, with deep enough
convection zones, may transport surface lithium to depths
where lithium can be destroyed, thereby gradually depleting the
lithium abundance in the photosphere. In metal-poor main sequence
stars, lithium abundances are observed to
decline with effective temperature for
$T_{\rm eff} < 5400$~K (see
\markcite{T94}Thorburn 1994 and references therein), presumably due to
the deeper convection zones for such stars.
The same effect is seen in metal-poor subgiants
(\markcite{PSB93}Pilachowski, Sneden, \& Booth 1993). 
However, among the hotter main-sequence dwarfs and subgiants, lithium
abundances in metal-poor stars show a near-constant
value of log~n(Li)~=~2.2, which is often referred to as the
``Spite Plateau", after its discovery by \markcite{Spite82}Spite \& Spite (1982).
There is now an extensive literature on lithium abundances in
metal-poor stars (see \markcite{SMFS93}Spite, Molaro, \& Spite 1993 and 
\markcite{R01a}Ryan et al.\ 2001
for additional observations and references).

Mass transfer (or mixing or pulsation-driven
mass loss, for that matter) should lead to depletion of surface lithium
abundances. 
Indeed, lithium is not found in many candidate
blue stragglers studied to date (\markcite{HM91}Hobbs \& Mathieu 1991;
\markcite{PG91}Pritchet \& Glaspey 1991; \markcite{GPS94}Glaspey,
Pritchet, \& Stetson 1994; \markcite{CLLGM}CLLGM). On the other hand, normal lithium
abundances have been found in some stars that are blue straggler
candidates, suggesting that they are either blue stragglers created
by an as-yet unidentified mechanism, or truly younger metal-poor stars
(\markcite{R01a}Ryan et al.\ 2001; \markcite{CLLGM}CLLGM). We therefore include
in Table~1 the lithium abundances for our program stars,
insofar as they have been determined by others.

Alas, the lithium abundance may not be entirely capable of distinguishing
a field metal-poor blue straggler star (low lithium) from a young
metal-poor dwarf (normal lithium). In the Hyades cluster, \markcite{BT86}
Boesgaard \& Trippico (1986) and \markcite{BB88} Boesgaard \& Budge (1988)
found that lithium abundances are very depleted in main sequence
stars with effective temperatures lying roughly in the 
range of 6400~K to 6800~K.
This ``lithium gap", if it arises during the evolution of normal
metal-poor main sequence stars, could lead us to confuse a young 
but otherwise normal metal-poor
dwarf with a metal-poor blue straggler. We discuss this further
in Section 5.2.

\subsection{Other Elemental Abundances}

\markcite{SPC2003}SPC2003 discovered strong enhancements of elements probably
created by $s$-process
nucleosynthesis in three stars that satisfy the color
conditions described below. These abundance anomalies speak
strongly to mass transfer, and also the evolutionary
state of the donor star, and hence we have searched the literature
to determine what is known about the abundances of two 
elements which are often associated with $s$-process nucleosynthesis,
strontium and barium. These abundances are also summarized in Table~1.

\subsection{Our Five New Program Stars}

We rely once more on a large sample of field stars to select our program
stars. \markcite{S81}Stetson (1981) primarily used proper motion criteria
to select possible high velocity stars from the SAO catalog, and obtained
a list of high-velocity early-type stars (\markcite{S91}Stetson 1991).
Based on high-resolution spectra, \markcite{GPS94}Glaspey,
Pritchet, \& Stetson (1994) found that these stars could be divided into
three broad groups of stars, one of which is relevant to metal-poor
field blue stragglers. Their Table~4 (part {\em iii}) includes six metal-poor
stars that are also deficient in lithium. One of these stars, SAO~80390
(BD+25~1981) was studied by \markcite{CLLGM}CLLGM, and found to be a possible radial
velocity variable. 
The other five stars constitute our new sample. 

In Table~1 we provide the basic photometric and chemical information
for these five stars, plus the ten blue straggler candidates from \markcite{CLLGM}CLLGM, and
the ten metal-poor ([Fe/H] $<$ $-0.6$) stars from \markcite{PS00}PS2000 for which 
\markcite{CLLGM}CLLGM could obtain satisfactory orbital solutions using their 
software and generally more
conservative criteria. The de-reddened $B-V$ values, effective
temperatures, and metallicities for the \markcite{PS00}PS2000 stars were taken
from that paper, while for BD+23~74, HD~8554, HD~109443, and HD~135449,
the values were taken from or determined using data taken from
\markcite{S91}Stetson (1991) and \markcite{GPS94}Glaspey et al.\ (1994). 
Values for the remaining
stars were determined following the same precepts discussed
by \markcite{XII}Carney et al.\ (1994).
In Figure~1 we show the color selection criterion for blue straggler
candidates. As discussed by \markcite{CLLGM}CLLGM, the solid line represents a second-order
polynomial fit to the de-reddened $B-V$ color indices of selected globular
cluster main sequence turn-offs as a function of metallicity. Because the
tail of the field star metallicity distribution extends
to lower metallicities than that of the globular clusters, we have also employed
the ``Y$^{2}$" isochrones (\markcite{YY2}Kim et al.\ 2002). This use
of the isochrones is far from rigorous since we ignore color shifts that may
be required to bring the isochrones into agreement with observed cluster
color-magnitude diagrams. Further, we have made the convenient, perhaps naive,
assumption that all globular clusters have the same age. We simply chose the
isochrones with enhanced abundances of 
the ``$\alpha$" elements, [$\alpha$/Fe] = +0.3,
and then chose the cluster ages to all be 11 Gyrs, represented by the
dashed line in Figure~1. This provides a reasonable match to the edge of
the color distributions of the field stars over a wide range of
metallicities, 
especially at the lowest metallicities, and a good 
match to the actual globular cluster turn-off distribution
near the peak of the halo metallicity distribution ([m/H] $\approx$ $-1.6$).
Stars blueward of {\em both} of these lines are considered blue straggler
candidates. Note that CS~$22873-139$ does not satisfy this criterion, and
we exclude it from the following discussions.

\begin{figure}
\epsscale{1}
\figurenum{1}
\plotone{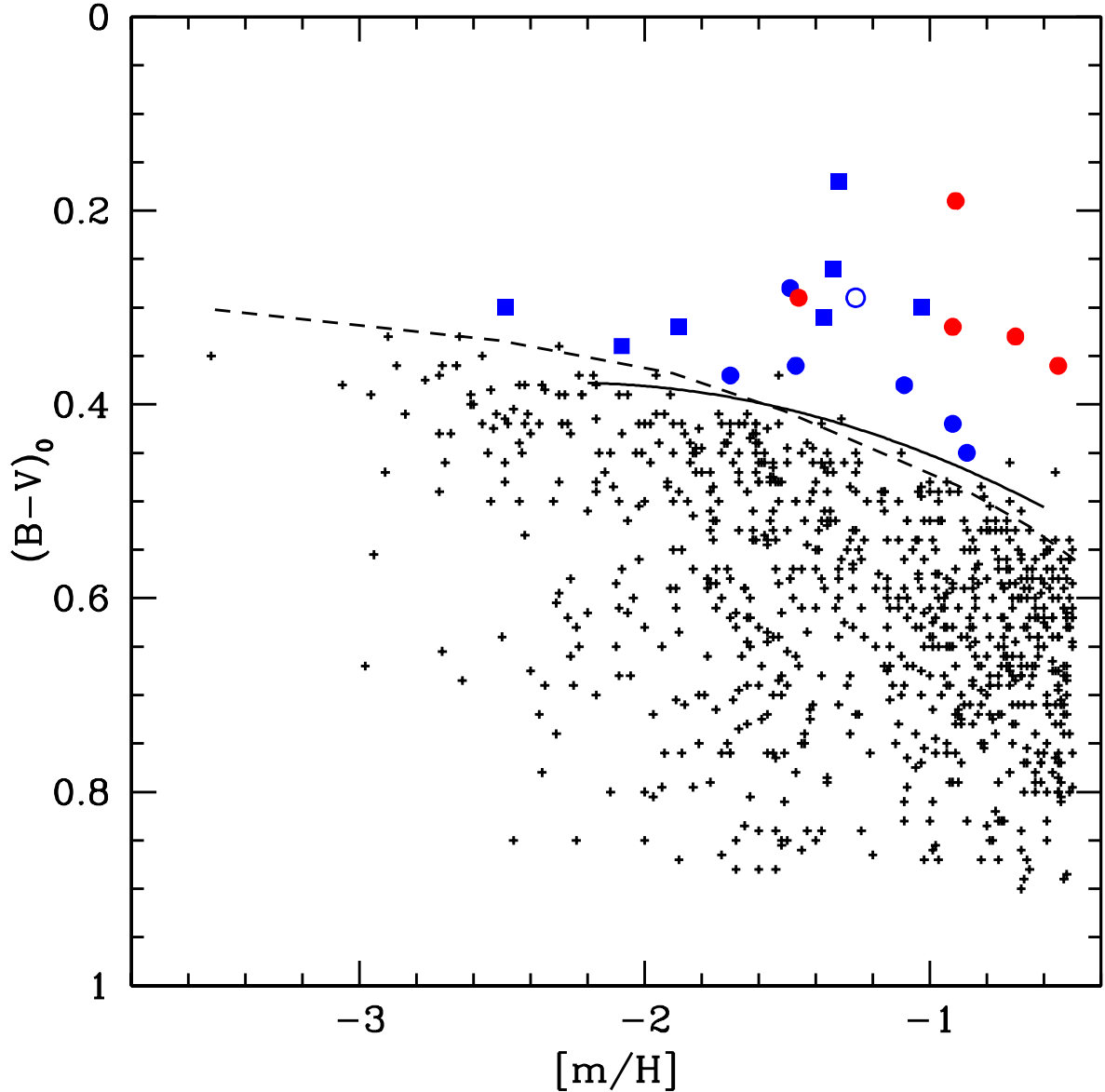}
\caption{De-reddened $B-V$ color indices are
plotted vs.\ metallicity for metal-poor field main sequence
stars (plus signs), and for the five new blue straggler candidates discussed
in this paper (red filled circles). Also shown are the six blue
straggler binaries with orbital solutions from Carney et al.\ (2001;
blue filled circles) and the seven blue straggler binaries with orbital
solutions and periods longer than 20 days from PS2000 (filled
squares). BD+25~1981,
which may be a binary star, is plotted as an open blue circle. The solid
line represents a fit to observed globular cluster main sequence
turn-off colors, and the dashed line is the theoretical equivalent
for clusters with identical ages, as discused in the text.
\label{fig1}}
\end{figure}

\placefigure{fig1}

\begin{table}
\dummytable \label{tab1}
\end{table}
\placetable{tab1}

The normal stars, not known to be blue stragglers, are plotted as
plus signs. However, compared to our earlier work (\markcite{CLLGM}CLLGM),
we have increased the number of normal stars for comparison.
We have completed and are preparing for 
publication an additional sample
of 470 metal-poor field stars selected from \markcite{R89}Ryan (1989) and
\markcite{RN91}Ryan \& Norris (1991). Three of those stars, 
G121-54, BD$-20$~3682,
and LTT~15049 (=HD~154578), plotted as plus signs
but bluer than the two dividing lines, are now blue straggler candidates. We will
report on their radial velocity behavior in a future paper.

We have also continued our
observations of the four ``constant velocity"
stars in our earlier paper, BD+72~94,
BD+40~1166, BD+25~1981, and HD~84937, and report the revised results as well.

\section{OBSERVATIONS}

The observations were made using the same procedures as described
by \markcite{CLLGM}CLLGM. We employed the
Center for Astrophysics (CfA) Digital Speedometers
\markcite{L85,L92}(Latham 1985, 1992) on 
the Multiple Mirror Telescope and 1.5-m Tillinghast Reflector at the
Whipple Observatory atop Mt. Hopkins, Arizona, and on the 1.5-m Wyeth
Reflector located in Harvard, Massachusetts.  The echelle spectrographs
were used with intensified photon-counting Reticon detectors and recorded
about 45~\AA\ of spectrum in a single order centered near 5187~\AA. The
resultant resolution
was about 8.5~\kms. Typical signal-to-noise values ranged from
from 7 to 50 per resolution element.

Radial velocities were measured from the observed spectra using the
one-dimensional correlation package {\bf rvsao} (\markcite{KM98}Kurtz
\& Mink 1998) running inside the IRAF\footnotemark \footnotetext{IRAF
(Image Reduction and Analysis Facility) is distributed by the National
Optical Astronomy Observatories, which are operated by the Association
of Universities for Research in Astronomy, Inc., under contract with
the National Science Foundation.} environment.  As before, we used
a grid of synthetic spectra 
calculated using the model atmospheres computed using R.\ L.\ Kurucz's code
ATLAS9. The details of the calculation of the models and the grid
of synthetic spectra were presented by \markcite{CLLGM}CLLGM.

For each of the five new blue straggler candidates we adopted a
template from the grid of synthetic spectra with the
effective temperature and metallicity nearest to the values for that
star reported in Table~1.  We used $\log g = 4.0$
throughout.  We computed correlations for all spectra of each star
using all the available rotationally broadened synthetic spectra
with the most appropriate temperature, gravity, and metallicity.
To determine the radial velocity, we chose the rotationally-broadened
template with the value of \vrotsini\
that gave the highest average
value for the peak of the correlation function. This
established the optimum set of parameters for the synthetic
spectrum template for each star.
The parameters
adopted for the final correlations are coded in the final column of
Table~1: effective temperature, log~g, [Fe/H], and \vrotsini.

Table~2 is an illustrative summary of the individual radial velocity measurements,
including the heliocentric Julian Day of mid-exposure,
and the velocity and internal error estimate returned by the IRAF task
{\bf rvsao}, both in \kms. The full table with all the velocity information
is available electronically.

\begin{table}
\dummytable \label{tab2}
\end{table}
\placetable{tab2}

\section{RESULTS}

\subsection{Radial Velocities}

We summarize our radial
velocity results in Table~3, including the total number
of velocities measured for each star, the span of our velocity
coverage (in days), the derived rotational 
velocity, $v_{\rm rot}$ sin~$i$, the mean
radial velocity, and the uncertainty of the mean velocity. Note that
for the binary stars, the mean radial velocity is not as appropriate
as the systemic velocity that emerges from the orbital solution. For
stars with orbital solutions, we therefore list here the systemic
velocity and its uncertainty. We include in Table~3 the measured rms external
error, $E$, and the mean internal error, $I$, of the velocity measurements
(see \markcite{KM98}Kurtz \& Mink 1998), and the ratio, $E/I$. Large 
values of $E/I$ ($\approx 1.5$
and above) are
suggestive of radial-velocity variability. We also employ the
probability, P($\chi^{2}$), that the $\chi^{2}$ value could be larger
than observed due to Gaussian errors for a star that actually has
constant velocity.  We employ the internal error estimate,
$\sigma_{\rm i,int}$, obtained from {\bf rvsao},
for each of $n$ exposures when calculating
$\chi^2$:
\begin{equation}
\label{eq:chisquared}
\chi^2 = \sum_{i=1}^n (\frac{x_i-<x>}{\sigma_{i,\rm int}})^2 .
\end{equation}
It is clear that at least four of our five new program stars
display radial velocity variability.

\begin{table}
\dummytable \label{tab3}
\end{table}
\placetable{tab3}

\subsection{Constant Velocity Stars}

HD~142575 does not appear to be a radial velocity variable, and we
show its velocity history in Figure~2. Table~3 shows that the probability
that the measured value of $\chi^{2}$ could arise by chance is large,
which is a strong indicator of radial velocity stability. The four
stars in our earlier work (\markcite{CLLGM}CLLGM) that appeared to have
constant radial velocities have rather small P($\chi^{2}$) values,
in particular for BD+25~1981. However, the ten new radial velocities
have resulted in a higher P($\chi^{2}$) value for this star than
in our earlier paper, when it was only 0.000002.
The rest of Figure~2 shows the velocity
histories of these four stars. No obvious signs of orbital motion are
seen, although BD+25~1981 remains an intriguing case.

\begin{figure}
\epsscale{1}
\figurenum{2}
\plotone{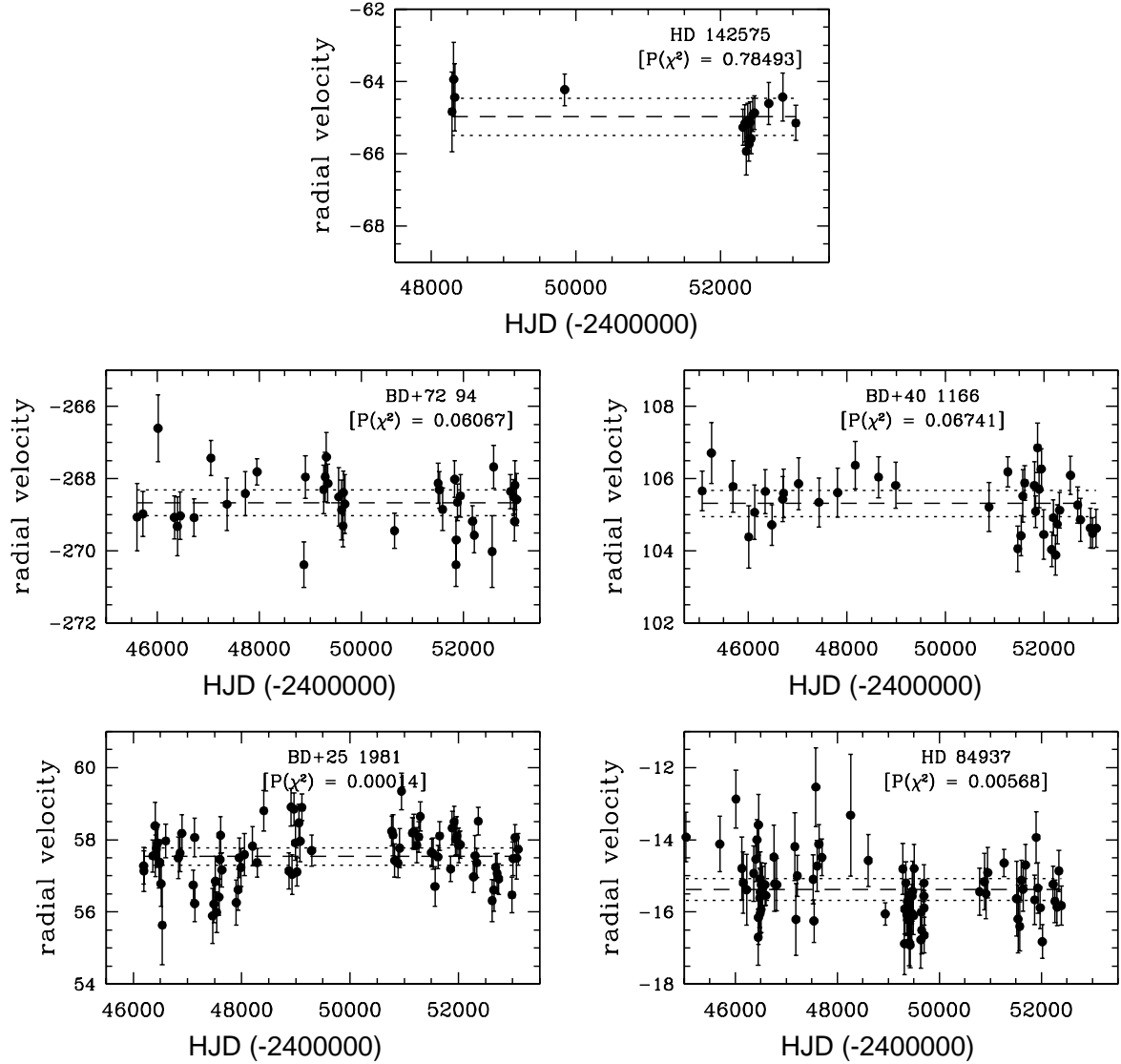}
\caption{The time histories of the measured radial
velocities for (a) HD~142575; (b) BD+72~94; (c) BD+40~1166; (d) BD~+25~1981;
and (e) HD~84937. \label{fig2}}
\end{figure}

\placefigure{fig2}

\subsection{Orbital Solutions}

Orbital solutions for four of the five new program stars are given
in Table~4, as well as calculated uncertainties. We include the
orbital period $P$ (in days), the systemic (mean) radial velocity, $\gamma$,
the orbital semi-amplitude, $K$, the orbital eccentricity, $e$, the
longitude of the ascending node, $\omega$, the time (in HJD)
of the periastron passage, the projected
semi-major axis of the primary star's orbit (in giga-meters), the
mass function, $f$($M$) (in units of a solar mass), the number of
observations (and the span in days of the observations), and, finally,
the probable error of the match of the observed velocities to the
orbital solution. Please note that our radial velocity observations
span several orbital cycles for all four stars.
The orbital solutions and the phased velocity data are displayed in
Figure~3.

\begin{figure}
\epsscale{1}
\figurenum{3}
\plotone{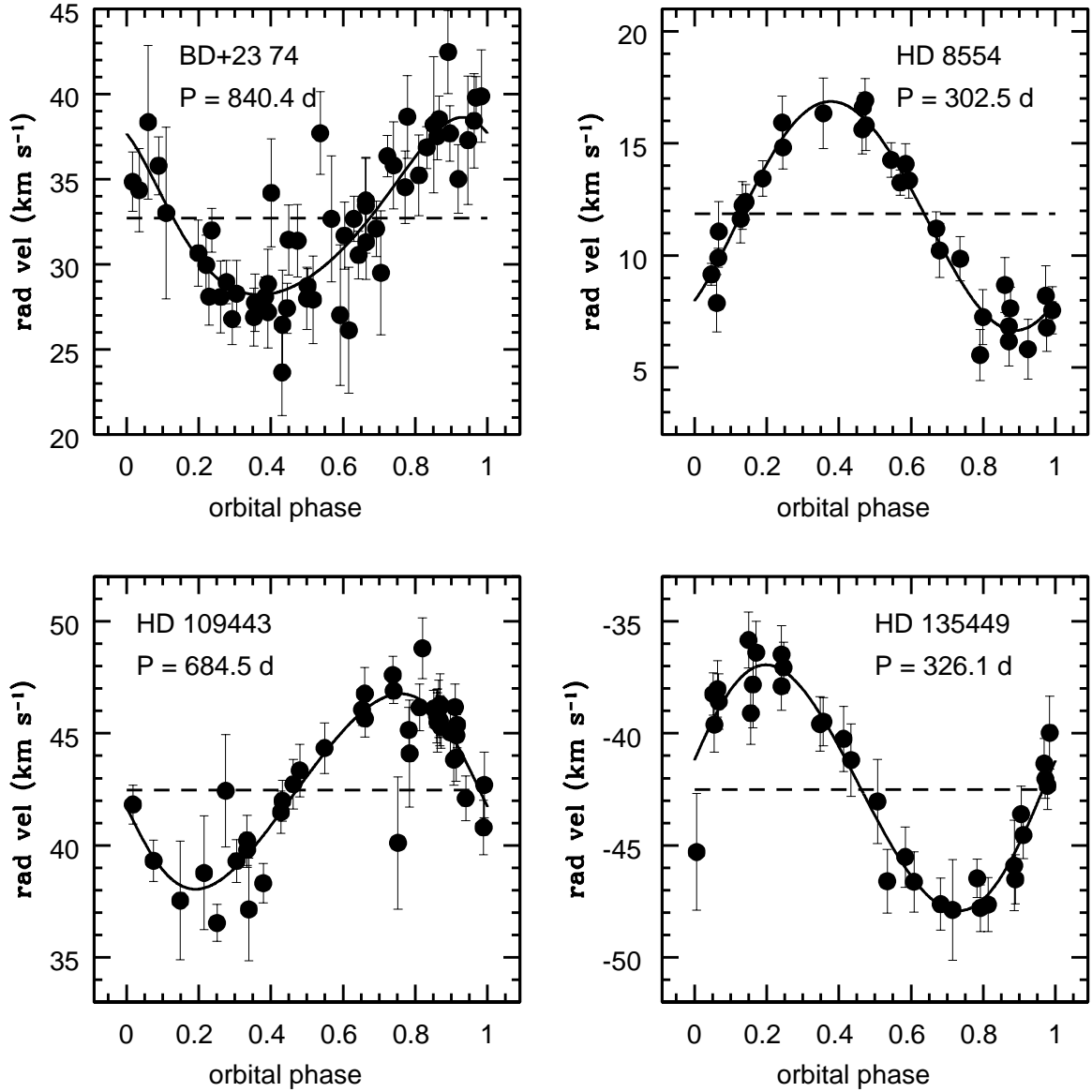}
\caption{The orbital solutions and observed
velocities for: (a) BD~$+23~74$; (b) HD~8554; (c) HD~109443;
and (d) HD~135449. \label{fig3}}
\end{figure}

\placefigure{fig3}

\begin{table}
\dummytable \label{tab4}
\end{table}
\placetable{tab4}

\subsection{Rotational Velocities} 

We have attempted to derive internally consistent values
for the
stellar rotational velocities, \vrotsini,
using our synthetic spectra for all of our program stars. We
have described above how we selected the optimum value
of \vrotsini\ for the synthetic spectrum to derive
the radial velocities. To provide the best estimate
for \vrotsini, we interpolated between the average peak correlation
values for all spectra using different values of
\vrotsini\ in the synthetic spectra.
We used quadratic fits of the correlation values vs.\
\vrotsini\ to select the final value. These are
given in Table~3.

We have tested the accuracy of our derived rotational velocities
by employing two sets of template spectra for each star.
The first choice of templates are those given in Table~1.
The second choice was made by the computer, with the template
selected independently on the basis of the height of the
peak of the correlation function for the fifteen stars with
templates listed in the Table. In six cases this resulted in
exactly the same adopted template. The other nine cases
enable us to determine how a change in temperature or
gravity alter our derived rotational velocity estimate. 
The greatest change was for HD~97916, where the computer's
preferred template was 750~K cooler and 0.5 dex lower in
gravity, and the $v_{\rm rot}$~sin~$i$ value changed by
only 1.8 \kms, down to 13.1 \kms. For BD+40~1166 and HD~84937,
the computer chose templates 250~K cooler and hotter, respectively,
than the values given in Table~1. Even here, where the derived
rotation velocities are comparable to or smaller than our
instrumental resolution, the changes were small, $-1.1$ and $+0.7$
\kms, respectively. For more rapidly rotating stars, the sensitivity
was even smaller. For BD+23~74, for example, a change of 1000~K
changed the derived $v_{\rm rot}$~sin~$i$ value by only 0.4 \kms.
The power of synthetic spectra and the $\chi^{2}$ fitting
technique's sensitivity to many weak
lines carrying the same information is demonstrated quite nicely
by these tests.

\section{DISCUSSION}

\subsection{Orbital Periods \& Eccentricities}

\markcite{PS00}PS2000 drew attention to the relatively long periods and low orbital
eccentricities of many of the binary stars they discovered.
In combination with the high binary frequency and small fraction of
double-lined systems, they argued that their results supported the
identification of many of these binary systems as blue stragglers
created by mass transfer, a model proposed originally by \markcite{McC64}McCrea (1964).

\markcite{CLLGM}CLLGM similarly drew attention to the unusual
orbital properties of the six field blue stragglers they studied and
the ten systems discovered by \markcite{PS00}PS2000 with [Fe/H] $\leq$\ $-0.6$ and
for which our own software resulted in what we considered compelling
orbital solutions. Neglecting the three \markcite{PS00}PS2000 systems with periods
of less than 20 days (and whose orbital elements may have been altered
by tidal interactions, independent of mass transfer), we found
that 10 of the 13 binaries had $e < 0.15$ and $<e> = 0.062 \pm 0.011$.
For all 13 systems, $e < 0.29$ and $<e> = 0.106 \pm 0.025$. The periods ranged
from 167 days up to roughly 1600 days. One should recall that the median
orbital eccentricity for normal main sequence, single-lined spectroscopic
binaries is about 0.37, independent of metallicity (\markcite{DM91}Duquennoy \& Mayor 1991;
\markcite{XVI}Latham et al.\ 2002). 

Mass transfer appears to be the best explanation for these orbital
characteristics, as discussed originally by \markcite{W86}Webbink (1986). Our earlier
paper also discussed stellar systems where mass transfer is likely to
have occurred, leaving an overabundance of elements synthesized
during the AGB evolutionary stage
deposited in the outer layers of the receptor. If we confine
our attention to the more metal-poor stars, there are three
classes of stars to consider. The CH stars are not luminous enough
to be AGB stars themselves, yet show abundances expected of such
stars, implying mass transfer has occurred and also that the companion
star is likely to be a white dwarf. 
\markcite{MW90}McClure \& Woodsworth (1990) found that
essentially all CH stars they studied are binaries, and those with
orbital solutions showed periods of 328 to 2954 days, and low orbital
eccentricities ($e < 0.18$; $<e> = 0.05$). The ``subgiant CH" stars
discovered by \markcite{B74}Bond (1974) may be descendents of some blue stragglers,
as noted by \markcite{LB91}Luck \& Bond (1991). 
\markcite{M97}McClure (1997) found that 9 of the 10
such stars under long-term radial velocity monitoring are binary systems
with long periods. He
obtained orbital solutions for six of the
sytems, finding orbital periods
of 878 to 4140 days, and small orbital eccentricities
($e < 0.13$; $<e> = 0.11 \pm 0.02$, $\sigma = 0.04$). 
Finally, the one well-studied dwarf
carbon star, G77-61, has been found to have a nearly circular orbit
and a moderately long period of 245 days (\markcite{DLADHM86}Dearborn et al.\ 1986). 

One of the primary purposes of the observations reported in this paper
was to test the idea that many blue stragglers are binary stars, and
that their binary orbital properties confirm this kinship with the
more obvious mass transfer systems described above. In Figure~4 we
update Figure~4 from \markcite{CLLGM}CLLGM, which showed the
orbital eccentricity vs.\ log~P for the 6 new binary systems reported therein,
and the 10 systems fom \markcite{PS00}PS2000 subjected to the same orbital solution
analyses. For comparison, we included the 156 orbital solutions for
cooler single-lined spectroscopic binaries from \markcite{XVI}Latham et al.\ (2002).
In this new Figure~4, we highlight the locations of the four new
orbital solutions from Table~4 using filled circles. 
(Filled squares are our results from \markcite{CLLGM}CLLGM; filled diamonds are
stars from \markcite{PS00}PS2000 with periods longer than 20 days, and
open diamonds are stars from \markcite{PS00}PS2000 with shorter periods.)
As before, the longer orbital period binary stars have unusually
low values of the orbital eccentricity.

\begin{figure}
\epsscale{1}
\figurenum{4}
\plotone{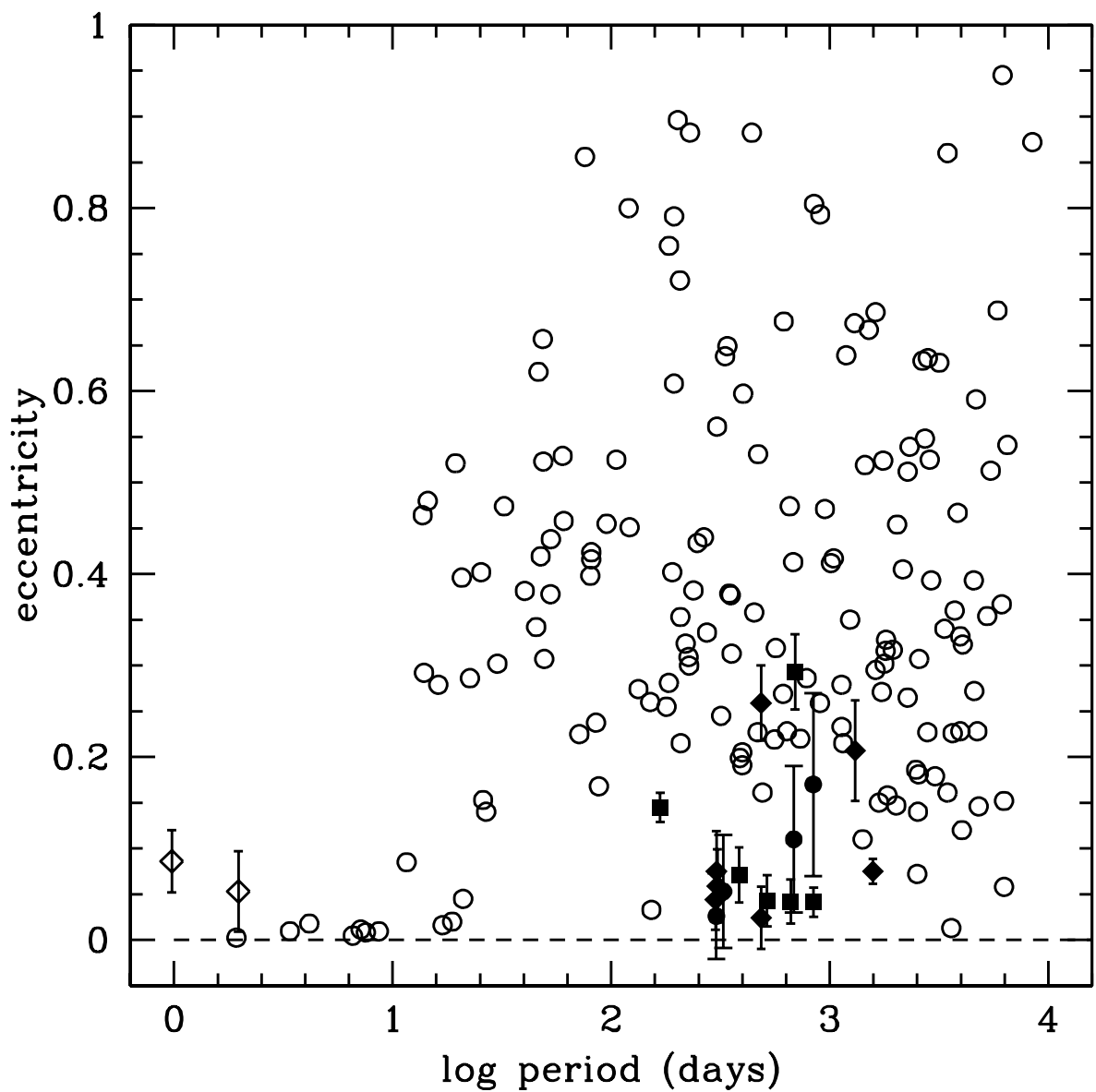}
\caption{The orbital eccentricity vs.\ log period
for 156 single-lined spectroscopic binaries from Latham et al.\
(2002) are plotted as open circles. The orbital solutions for the
blue stragglers shown in Figure~1 are also shown. 
Filled circles are the new orbital solutions from Table~4, filled
squares are our orbital solutions from CLLGM, filled diamonds
are stars from PS2000 with periods longer than 20 days, and
open squares are stars from PS2000 with shorter periods. \label{fig4}}
\end{figure}

\placefigure{fig4}

In Figure~5, we take a different perspective,
and consider the distribution of orbital eccentricities for
stars in Figure~4 with periods longer than 20 days. Aside from the
absence of high orbital eccentricities among the blue
stragglers, we note 
that the orbital eccentricities cluster near zero but have
a tail to higher values.
This might indicate two
different types of mass transfer and orbital evolution, despite the
similarity of the orbital periods. Alternately, it may reflect the original
distribution of orbital eccentricities since binary systems with
very high eccentricities evolve more slowly, and so are more likely
to preserve a higher-than-average orbital eccentricity.

\begin{figure}
\epsscale{1}
\figurenum{5}
\plotone{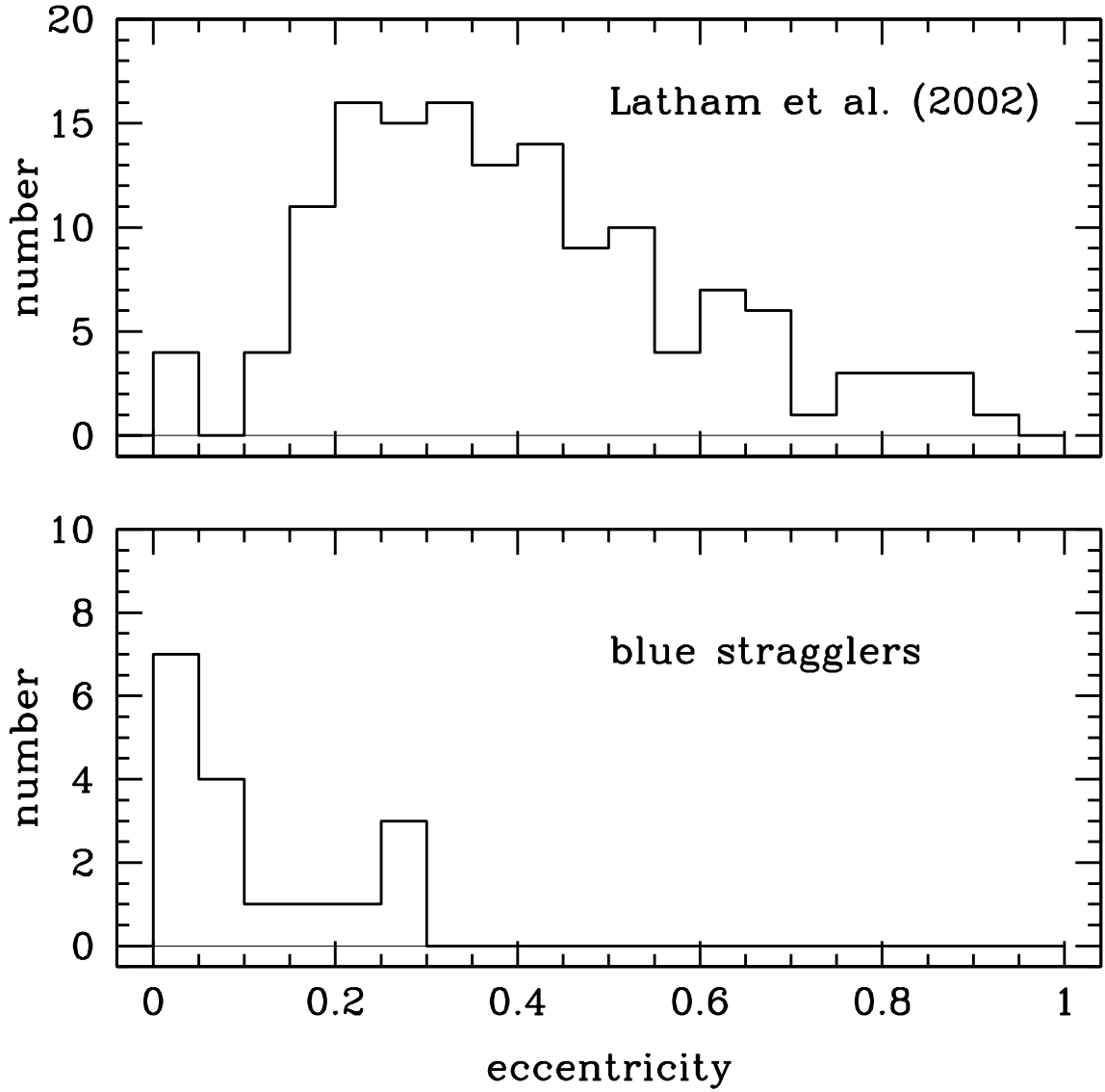}
\caption{The distributions of the orbital
eccentricities of the stars shown in Figure~4. \label{fig5}}
\end{figure}

\placefigure{fig5}

\subsection{Lithium}

In Figure~6 we show the ``lithium gap" seen among some members of
the Hyades, using the results of \markcite{CCCD84}Cayrel et al.\ (1984),
\markcite{BT86}Boesgaard \& Trippico (1986), and
\markcite{BB88}Boesgaard \& Budge (1988). 
Open circles are measurements and ``{\sf v}" symbols
represent upper limits. The gap is easily seen between
effective temperatures of about 6400~K and 6800~K. 
The normal lithium abundance for metal-poor dwarfs hotter
than about 5400~K (the ``Spite plateau").
is shown as a dashed line. That value is lower than the peak
of the Hyades dwarfs and 
is in itself a long-standing puzzle,
but beyond the point of this paper. 
We also show the stars from Table~1, with filled circles
for stars that have measured lithium abundances, and which, it
turns out, are all also stars with constant velocities.
None of the binary stars in Table~1 have detectable
lithium lines and hence the abundances are only upper limits.
These stars are plotted as downward-pointing arrows. Note
that for HD~142575, 
\markcite{GPS94}Glaspey et al.\ (1994) derived $T_{\rm eff}$ = 6550~K,
and \markcite{JFB2000}Fulbright (2000) adopted $T_{\rm eff}$ = 6500~K, but we
have adopted the photometric temperature of 6700~K to maintain
consistency with the derivation of the temperature for all of
our program stars. We have adjusted the lithium abundance derived
for the star by \markcite{JFB2000}Fulbright (2000) as well to account for
the change in temperature: log~n(Li) rises from 1.45 to 1.58.
(Note as well, however, that \markcite{GPS94}Glaspey et al.\ 1994 argued that
log~n(Li) $<$ 0.7.)

\begin{figure}
\epsscale{1}
\figurenum{6}
\plotone{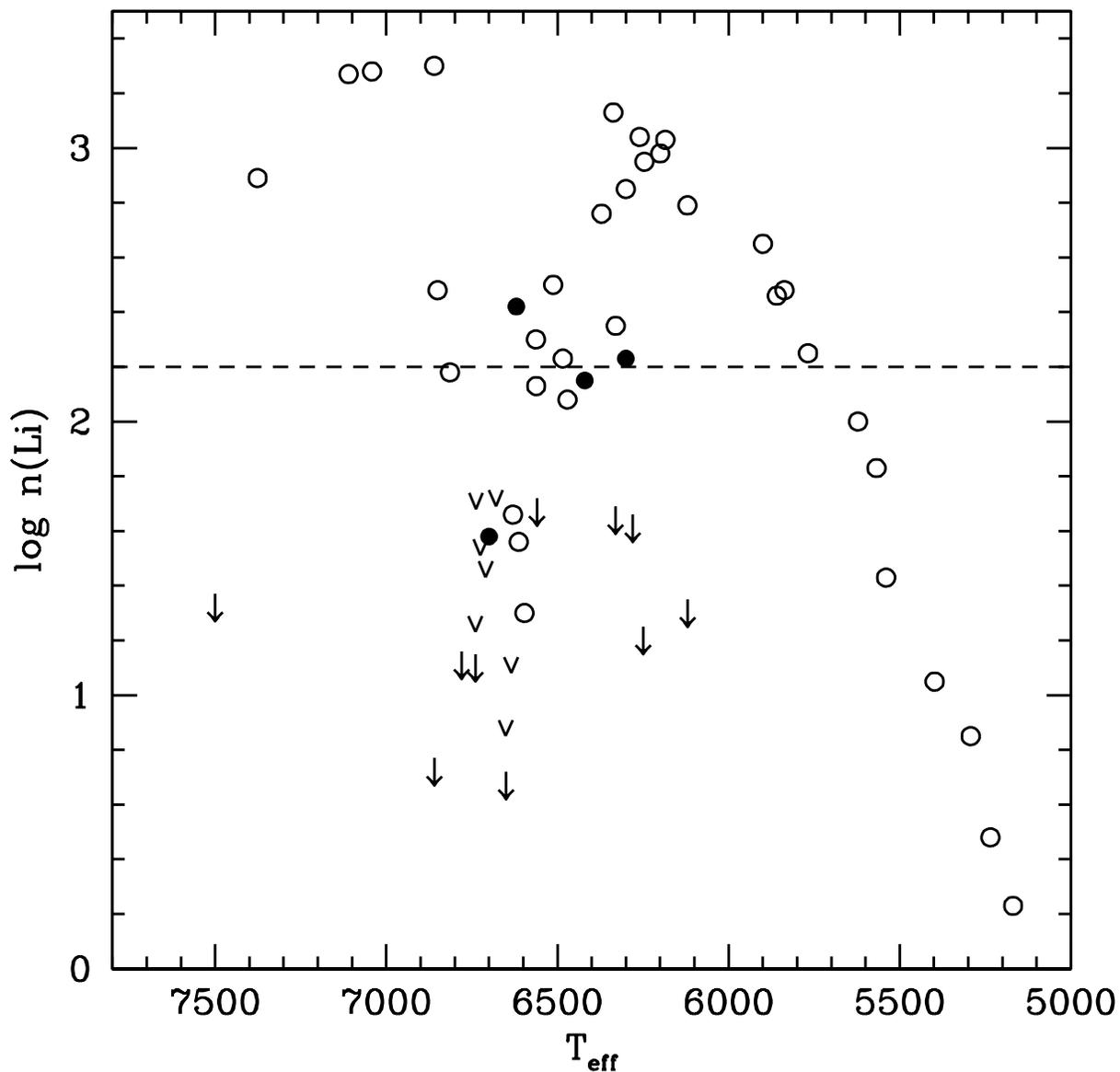}
\caption{The derived abundances of lithium
in Hyades dwarfs are shown as open circles, taken from
Cayrel et al.\ (1984), 
Boesgaard \& Trippico (1986), and
Boesgaard \& Budge (1988). Stars with only upper limits
are plotted as ``{\sf v}".
The lithium abundances of the
four stars discussed in this paper are shown as filled circles,
and upper limits are signified by arrows, 
as summarized in Table~1.
The dashed line represents the ``Spite plateau" of
lithium abundances for metal-poor main sequence stars with
6200 $>$ T$_{\rm eff}$ $>$ 5400~K. \label{fig6}}
\end{figure}

\placefigure{fig6}

The first interesting point is that three of the metal-poor single
stars have normal lithium abundances, at least for
metal-poor stars. Two of the stars,
HD~84937 and CS~$22873-139$, are very close to our adopted
color limits used to identify blue stragglers. Small changes
in the measured $B-V$ color index or in the reddening estimate
could easily shift either star into or out of the blue straggler
candidate domain. However, BD+72~94
is considerably bluer and hotter, and in the middle of the Hyades
lithium gap, yet has a lithium abundance consistent
with the Spite plateau.
Conclusions from such a small sample  may not be secure,
but the results suggest the lithium gap may not exist for
single metal-poor stars, or, perhaps, it is shifted to
higher temperatures than the stars in our sample. A survey of
the metal-poor single stars identified by \markcite{PS00}PS2000 would answer
this question.

Related to this point, HD~142575, which appears to be single and slightly
hotter than the other three single stars,
has a reduced lithium abundance according to \markcite{JFB2000}Fulbright (2000).
Is this star distinguishable in any way from the three
constant velocity, normal-lithium stars? If our photometric
temperature estimate is correct, the star is slightly hotter
than the other single stars, and so might be in the metal-poor 
equivalent to the Hyades lithium gap.
Table~3 may provide another
clue: HD~142575 has a relatively large value for \vrotsini. Is rotation
a factor in either the formation of the lithium gap or the
depletion of lithium more generally? We need more observations
of lithium absorption line strengths in blue metal-poor stars,
especially those studied by \markcite{PS00}PS2000.

We conclude that our use of low lithium abundances as a selection
criterion is supported by the (admittedly limited) data available.
We note also that the relation between blue stragglers
and lithium depletion has been discussed extensively 
by \markcite{R01a}Ryan et al.\ (2001).

\subsection{Rotation}

\markcite{RGKBK}Ryan et al.\ (2002) have noted the modest
but significant levels of rotation in three of the blue stragglers
discussed in our previous paper (\markcite{CLLGM}CLLGM). For 
BD+51~1817, G66-30, and G202-65, they found $v_{\rm rot}$~sin~$i$
values of $7.6 \pm 0.3$, $5.5 \pm 0.6$,and $8.3 \pm 0.4$ \kms,
all of which are larger than values derived for cooler metal-poor
dwarfs with normal lithium abundances. (Note that our method
obtains somewhat higher values of 9.0, 8.1, and 
11.6 \kms, respectively.)
\markcite{RGKBK}Ryan et al.\ (2002)
suggested that the likely origin of the rotation was the mass transfer
process itself, and discussed the mechanisms by which such a process
would lead to lithium depletion. These include mass transfer-induced
deep mixing or transfer of lithium-depleted material from the
donor star. Regarding the latter point, 
\markcite{PSB93}Pilachowski et al.\ (1993)
found that surface lithium depletion in post-main sequence metal-poor
stars begins before the stars reach the base of the red giant branch,
so mass transfer from a more highly evolved red giant would deposit
lithium-depleted material on the recipient main sequence star.

We explore here the available rotational velocity information
for the stars of Table~3. As described above, we are able to
estimate \vrotsini\ using our grid of synthetic spectra.
Because different approaches appear to derive different
values of \vrotsini, we proceed differentially by comparing
\vrotsini\ values of constant velocity and binary stars only
within a particular methodology. Thus we make comparisons
only between stars we have studied, or among the stars studied
by \markcite{PS00}PS2000, but we do not intermingle those results.
For the stars for which
we have estimated \vrotsini, 
we must first
explore the results for a ``control sample". In Figure~7
we show the histograms of the derived \vrotsini\ values using data
taken from \markcite{XVI}Latham et al.\ (2002). We include 
only single stars judged to be main sequence dwarfs, and single-lined
spectroscopic binary stars with periods exceeding 25 days.
Tidal circularization effects are clearly present for metal-poor
dwarfs with orbital periods of 20 days or less, but should not
be significant for the longer period systems we consider here.
None of these three different groups were found to have
different distributions of \vrotsini. However, metallicity
does seem to have a small influence on the \vrotsini\ values
we derive, as may be seen in Figure~7. 
We do not claim that Figure~7 represents the true distribution
of rotational velocities of metal-poor dwarfs because our instrumental
resolution ($\approx 8.5$ \kms) prevents accurate determinations
for values much smaller than that.
On the other hand, we are confident of our
results for rotational velocities at or above that
of our instrumental resolution.
The fraction of stars with \vrotsini\ $\geq$ 8.0 \kms\
in the combined sample is only 4.0\% (7.7\% for the more metal-poor
stars and 0.3\% for the more metal-rich stars). Note that we do
not believe these results imply that metal-poor dwarfs rotate somewhat
faster than metal-rich dwarfs. The differences may be artificial in the
sense that the synthetic spectrum templates may introduce metallicity-dependent
differences in the derived rotational velocities.

\begin{figure}
\epsscale{1}
\figurenum{7}
\plotone{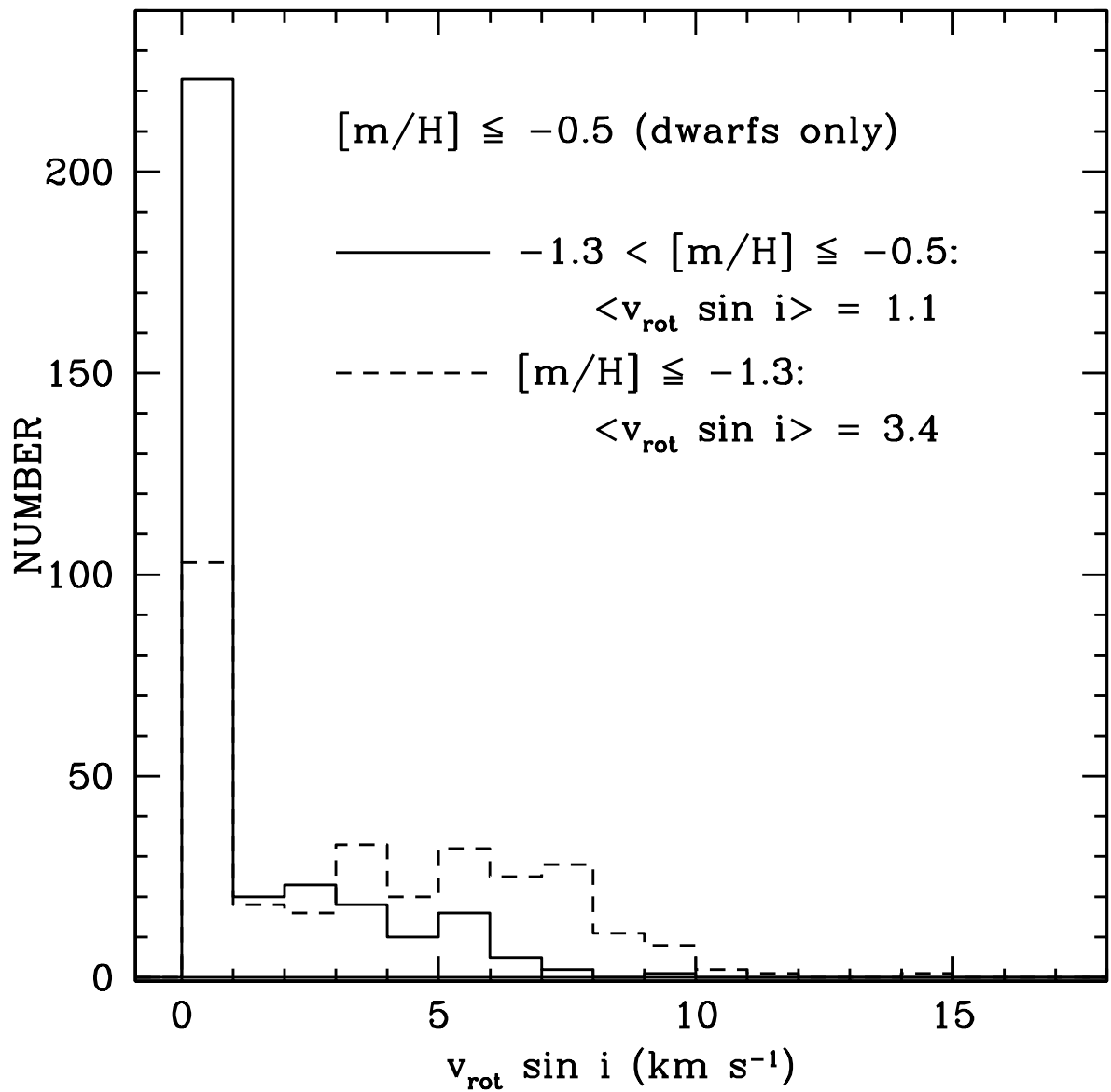}
\caption{The distribution of \vrotsini\
values for metal-poor main sequence stars, taken from
Latham et al.\ (2002). Recall that our instrumental
resolution is 8.5 \kms, so values much smaller
than that should not be considered reliable. Very
few metal-poor dwarfs show signs of rotation at or
above the level of our instrumental resolution. \label{fig7}}
\end{figure}

\placefigure{fig7}

In Figure~8 we show the rotational velocities we have derived
(open and filled circles) for the stars of Table~3, plus \vrotsini\
values from \markcite{PS00}PS2000 (open and filled squares). While the observations
of \markcite{PS00}PS2000 were obtained using a system with velocity resolution
comparable to ours, they derived \vrotsini\ values using a
different method than have we, which is why we distinguish
the two sets of results graphically. The key question is whether
mass transfer is a major cause of the blue straggler phenomenon
and whether evidence for it exists in higher rotational velocities
for the binary systems. Because tidal interactions will spin
up short-period binary systems, independent of mass transfer,
we have excluded binary systems with orbital periods of less
than 20 days from the Figure or from further consideration here.

\begin{figure}
\epsscale{1}
\figurenum{8}
\plotone{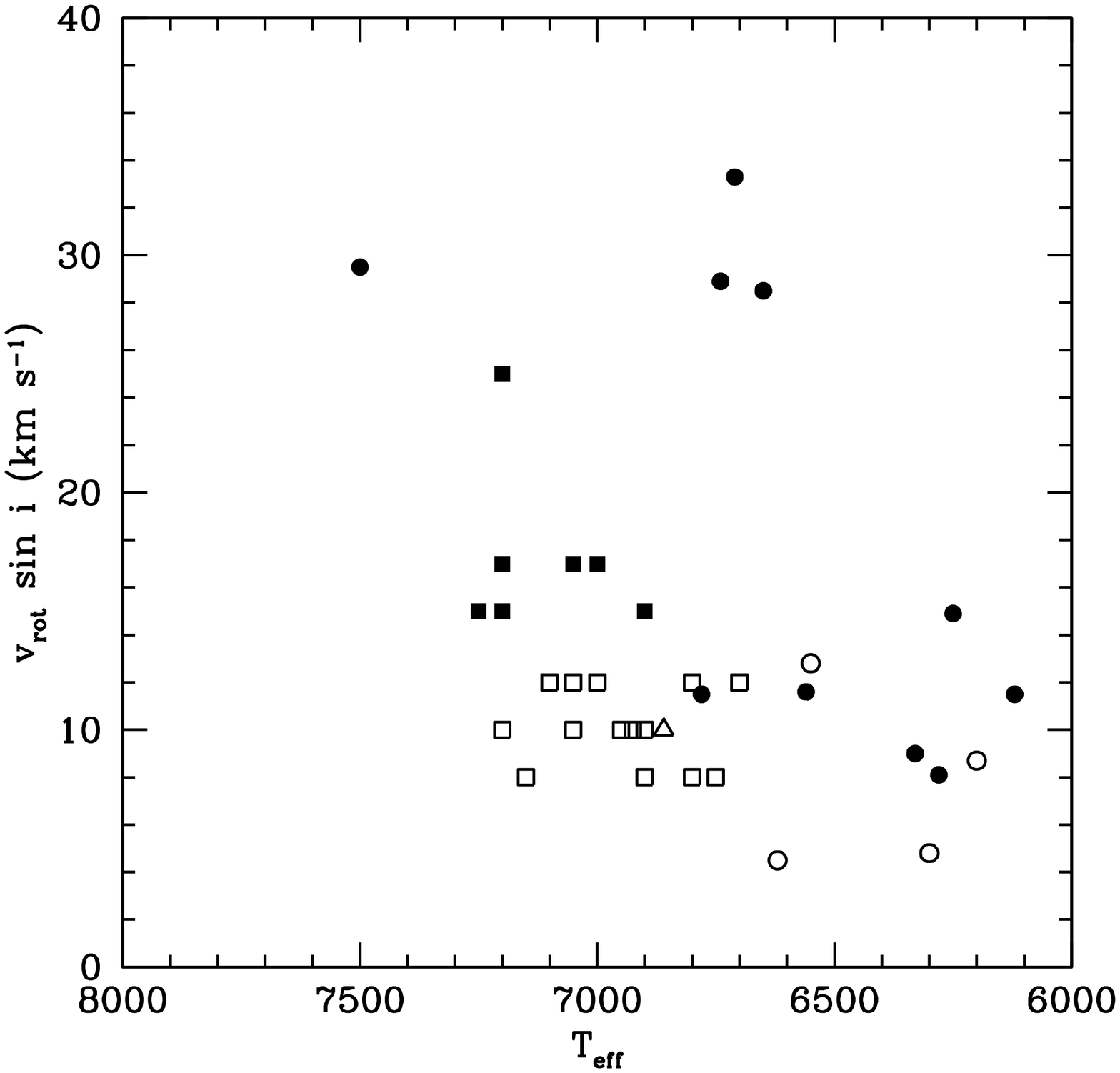}
\caption{The derived \vrotsini\ values
for single stars hotter than comparable metallicity main
sequence turn-off stars taken from PS2000 (open 
squares), as well as binary blue stragglers from PS2000
(filled squares) plus constant velocity
stars (open circles) from  Carney et al.\ (2001) and this paper
and binaries (filled circles).
The possible velocity variable BD+25~1981
is plotted as a triangle. \label{fig8}}
\end{figure}

\placefigure{fig8}

We consider the results from \markcite{PS00}PS2000 first. There are 14
stars in their study that appear to have constant velocity
and have [Fe/H] $\leq$\ $-0.5$. Recall that they have interpreted
(correctly, we believe) these stars to be bona fide young
metal-poor main sequence stars. Note that all of them
have \vrotsini\ $\leq$\ 12 \kms, and $<$\vrotsini$>$ =
10.1 \kms\ ($\sigma$ = 1.7 \kms). CS~$22873-139$,
which is a binary system but which we do not consider to
be a blue straggler candidate, has \vrotsini\ = 10 \kms.
[We have assigned CS~$22960-058$ a temperature of 6925~K
rather than 6900~K so it would not be overplotted on another star.]
If we restrict the sample to those stars from \markcite{PS00}PS2000
for which we could obtain an independent orbital solution
that agrees with their results (\markcite{CLLGM}CLLGM, Table~4), we find that
none of the seven binary stars with orbital periods longer than 20 days
have \vrotsini\ values as small as the largest value found for
the single stars, and that $<$\vrotsini$>$ = 17.3 \kms\
($\sigma$ = 3.5 \kms). [Here we have altered the
temperature for CS~$29518-039$ from 7050~K to 7000~K
to inhibit it being overplotted on another star.]
If we further consider the 11 more
binary stars for which \markcite{PS00}PS2000 presented only one orbital
solution, but which are not plotted in Figure~8,
only two have \vrotsini\ values of 12 \kms\ or less,
while the remaining nine stars have values reaching as high
as 160 \kms. 

Our own results, from our previous work (\markcite{CLLGM}CLLGM) and this paper,
confirm this picture. We consider
BD+25~1981 an uncertain case and have plotted it in Figure~8
as an open triangle. The other four constant velocity
stars (BD+72~94, BD+40~1166, HD~84937, HD~142575) have
$<$\vrotsini$>$ = 7.8 \kms ($\sigma$ = 3.8 \kms), while
the ten binary stars in Table~3 have $<$\vrotsini$>$ = 18.7 \kms
($\sigma$ = 10.0 \kms). 

In summary, our own studies and the results of \markcite{PS00}PS2000 show that
blue straggler candidates which are binaries show significantly
enhanced surface rotation compared to constant velocity
stars. Because the orbital periods imply separations far too
great for tidal interactions to {\em currently} have any significant
effects on the stars' rotational periods, we believe the enhanced
rotation indicates mass transfer has occurred and is the cause
of the spin-up.
The mass transfer model predicts
that the originally more massive star transfers mass to the lower
mass secondary, diminishing the orbital period and separation,
until the original secondary star becomes the more massive component.
At that point, the stars begin the separate, but mass transfer has
transformed orbital angular momentum into rotational angular momentum,
and tidal interactions or some mechanism has reduced the orbital
eccentricities.

\subsection{AGB Signatures: Carbon and $s$-process Elements}

As noted above, \markcite{LB91}Luck \& Bond (1991) suggested that some blue
stragglers may be progenitors of the subgiant CH stars.
Until recently, there has been little information regarding
abundances of carbon and $s$-process abundances in blue
stragglers that would indicate mass transfer from an AGB star,
but such data are now beginning to appear, and so we have
summarized in Table~1
the abundances of the $s$-process elements strontium and barium.
The available data for most candidate
blue straggler stars are consistent
with the normal abundances of these elements in metal-poor
main sequence and red giant branch stars (see 
\markcite{McW97}McWilliam 1997). Recently, \markcite{SPC2003}SPC2003
have found strong enhancements of these elements in three
of the blue metal-poor stars discussed previously by \markcite{PS00}PS2000.
Thus for these stars
the evidence favors 
mass transfer from an evolving AGB star, supporting
the hypothesis of \markcite{LB91}Luck \& Bond (1991).

If mass transfer from a highly-evolved AGB star is responsible
for the $s$-process enhancements, whereas the stars without
such enhancements experienced mass transfer from a normal red giant,
we might expect to see some differences in the orbital
properties of the two samples. Mass transfer from an AGB star
could imply, for example, that mass transfer did not occur during
the preceding red giant evolutionary phase, and, therefore,
the original orbital periods of binary systems destined to
experience mass transfer from an AGB star should be longer than
those which undergo mass transfer from a first ascent red giant.
One might expect, then, that blue stragglers showing signs of
AGB mass transfer ($s$-process enhancements)
should have longer minimum orbital periods
than those that (presumably) arose from mass transfer from
a red giant (no such enhancements). 

Qualitatively, Table~1 confirms this view. Excluding CS~$22956-028$, the eight
stars with normal strontium and barium abundances have orbital periods
as low as 167 days, and none exceed 500 days. The six subgiant CH stars
with orbital solutions obtained by 
\markcite{M97}McClure (1997) have periods only as
short as 878 days, and extend up to 4140 days. The eight (giant) CH stars
studied by \markcite{MW90}McClure \& Woodsworth (1990) have orbital periods ranging from
328 and 2954 days, roughly consistent with this simple picture. CS~$22956-028$ is
consistent with this trend, with an orbital period of 1307 days 
(\markcite{SPC2003}SPC2003).
Its orbital eccentricity is also on the high side (0.21) for the ensemble
of field halo blue stragglers, as Figure~4 shows. 

However, the other two stars with $s$-process enhancements, CS~$29497-030$ and
CS~$29509-027$, show rather short periods, according to 
\markcite{SPC2003}SPC2003, with 342 days
(and e $\approx 0$) and 194 days (and e $\approx 0.15$), respectively. We
have not included these stars in Table~1 because our orbital solutions
using the data available to \markcite{SPC2003}SPC2003 indicated that alternative periods and
eccentricities were possible, although less likely. Preston (private
communication) has continued to observe these stars and these orbital
periods are increasingly robust. In particular, we have been
able to confirm to our own satisfaction the orbital
period for CS~$29497-030$\footnote{The confirmation of the
short orbital period for CS~$29497-030$ has 
cost the first author payment of a wager
with Dr.\ Preston that longer periods would be preferred.}. Additional
velocity data are needed, however, for CS~$29509-027$.

We are therefore confronted with the results that mass transfer from
red giant and AGB stars yield similar ranges in orbital periods and
eccentricities. We can only speculate that diversity in
the masses of the original primary stars and their original orbital periods
contribute to this splendid if confusing variety.

\subsection{The Mass Transfer Process and Its Outcome}

A point stressed by \markcite{MW90}McClure \& Woodsworth (1990) and
\markcite{M97}McClure (1997) was that the mass functions of the CH
and subgiant CH stars, respectively, were small and
indicative of a uniform secondary mass, roughly
that of a white dwarf with 0.6\msun. We undertook
a similar analysis in our previous paper, which
we revisit here. 

\subsubsection{The Mass Functions}

We begin by comparing the distribution
of the mass functions of the blue stragglers and
the somewhat lower masses of the primary stars in the single-lined 
spectroscopic binaries of \markcite{XVI}Latham et al.\ (2002)
for stars with orbital periods longer than 20 days.
The mass function for a spectroscopic binary is defined to be
\begin{equation}
\label{eq:massfunction}
f(M) = sin^3 i \frac{M^{3}_{2}}{(M_{1} + M_{2})^{2}},
\end{equation}
where the numerical value of $f(M)$ is derived from the parameters
of the orbital solution. 

Figure~9 shows that the two mass function distributions are
quite different, with the steeper one for the blue stragglers
indicative of a narrower range of secondary masses, as in
the cases of the CH and subgiant CH stars.

\begin{figure}
\epsscale{1}
\figurenum{9}
\plotone{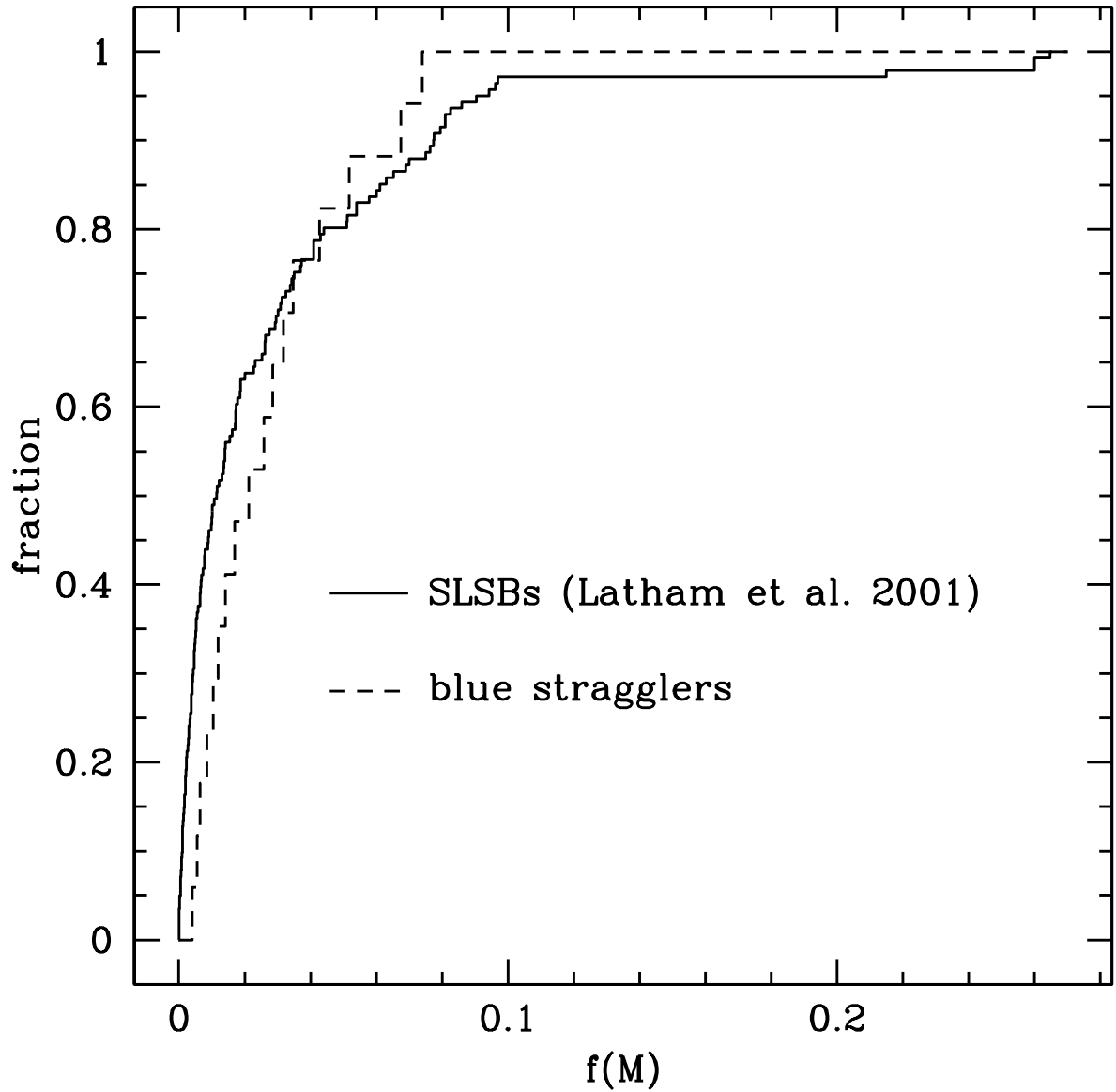}
\caption{The cumulative fractional distribution
of the mass functions of the single-lined spectroscopic
binaries from Latham et al.\ (2001) and the blue stragglers
in Table~5. \label{fig9}}
\end{figure}

\placefigure{fig9}

\subsubsection{Estimating the Secondary Masses}

So what is that secondary mass? We must begin with an estimate
of the masses of the primary stars, then exploit the mass functions
to get some idea of the secondary masses of the ensemble. As
we have noted, the lack of a detectable secondary spectrum in any
of the blue straggler binary systems is suggestive of underluminous
secondaries, and we explore here the idea that all the secondaries
are stellar remnants with similar masses.

In \markcite{CLLGM}CLLGM, we estimated the mass of each primary star
using the derived effective temperatures and chemical
compositions, plus the isochrones of \markcite{G2000}Girardi et al.\
(2000). The isochrone ages were relatively young,
generally 1.7 to 2.0 Gyrs, to accomodate the high temperatures
and apparent high masses of the primaries. The isochrones were computed
using elemental abundances consistent with solar values and,
to allow for enhanced abundances of the ``$\alpha$" elements
([$\alpha$/Fe]~$ = +0.3$), we employed 
the scaling relation of \markcite{SCS93}Salaris, 
Chieffi, \& Straniero (1993) to
calculate $Z_{\rm eff}$ for each primary. Here we exploit
the ``Yonsei-Yale" (Y$^{2}$) isochrones of \markcite{y2}Yi, Kim, \& Demarque (2003),
which were computed using $\alpha$-enhanced abundances and
somewhat more appropriate (and variable) helium mass fractions.
We used isochrones with an age of 2.0 Gyrs in all cases.
Table~5 shows the results. We have also compared these new
primary mass estimates with those computed using the 
\markcite{G2000}Girardi et al.\ (2000)
isochrones. The maximum difference in derived primary mass was 0.04 \msun,
and the mean difference, in the sense of \markcite{y2}Y$^{2}$ minus 
\markcite{G2000}Girardi et al. (2000),
was $0.005 \pm 0.005$ \msun, with $\sigma = 0.018$ \msun. 

\begin{table}
\dummytable \label{tab5}
\end{table}
\placetable{tab5} 

From the estimated value of $M_{1}$, we can estimate the \underline{minimum}
mass of the undetected companion; i.e.\ for $ \sin i = 1$, using
a revised version of Equation \ref{eq:massfunction}:
\begin{equation} \label{eq:minimummass} 
M_2 \sin i = f^{1/3}(M) \times (M_1 + M_2)^{2/3}.
\end{equation}
These
are also provided in Table~5. All of the minimum masses lie
below the expected mass of a metal-poor white dwarf, $\approx 0.55$\msun.
They also lie far below those expected for a metal-poor
main sequence turn-off star ($\approx 0.8$\msun) or that
of a neutron star
($\approx 1.4$\msun). The mean difference in secondary masses derived
using the two sets of isochrones is only $0.002 \pm 0.004$ \msun,
with $\sigma = 0.016$ \msun.

If all the secondaries have roughly the same mass, is it closer
to that of a white dwarf or a neutron star? 
If we {\it assume}
that all the secondaries have the same mass of $0.55 \msun$, and
calculate the orbital inclinations that this would imply, we obtain the
$\sin i$ values listed in Table~5.  The average $\sin i$ for
the 17 stars in Table~5 is $0.69 \pm 0.05$ ($\sigma = 0.19$),
which is not far from
the value of $\pi/4 = 0.785$ expected for randomly oriented
orbits. (Use of the isochrones of \markcite{G2000}Girardi et al.\ 2000 leads to
an identical result.)
Increasing the assumed secondary mass to 0.65 \msun,
results in $<\sin i> = 0.61 \pm 0.04$ ($\sigma = 0.17$),
which is inconsistent with the expectation by
a significant amount, although we must recognize that our
results suffer some observational bias in that very low
values of sin~$i$ result in smaller observed radial velocity
variations and such systems are more likely to elude detection.
If we further assume the secondaries
are all neutron stars, with $M = 1.4$\msun, we find
$<\sin i> = 0.37 \pm 0.03$ ($\sigma = 0.11$), and we 
rule out that class of secondaries, at least in the general case.

We conclude that our orbital solutions are consistent with all the
unseen companions in our low-eccentricity long-period orbits being
white dwarfs viewed at random orbital orientations, which supports the
mass-transfer model for the formation of field blue stragglers.  If
all the secondaries are white dwarfs with the same mass, then that
mass can not be much smaller than $0.50 \msun$, because the three
largest minimum masses are $0.55 \pm 0.06 \msun$ (CS~$22956-028$), $0.52
\pm 0.06 \msun$ (CS~$29518-039$), and $0.50 \pm 0.01 \msun$ (G202-65),
where the quoted errors include only the contribution from the mass
functions and not from possible systematic errors in
our estimates of the primary masses. 
On the other hand, the secondary masses can not be larger than
about $0.65 \msun$, because then the average value for the orbital
inclinations would become seriously inconsistent with random orientations of the
orbits.  

\subsubsection{Stable or Unstable Mass Transfer?}

\markcite{RGKBK}Ryan et al.\ (2002) have discussed the mechanism
for the mass transfer that might have increased the
rotational velocities of the stars we see now as
blue stragglers. One of the questions they raised was whether
the mass transfer mechanism could have involved stable
Roche lobe overflow or whether the mass transfer process
was unstable. Sudden mass loss via a supernova, for example,
would be such an unstable process, although we have essentially ruled
out neutron star companions among this sample of field
blue stragglers. 

\markcite{rappaport95}Rappaport et al.\ (1995) modelled the results
of Roche lobe overflow.
They considered the evolution of binaries undergoing
stable mass transfer, but for the case where the recipient is a
neutron star.  The same physics should also apply to systems where the
recipient is a main-sequence star.
The essence of the model is that short-period systems are so close
that mass transfer begins early, when the expanding red giant core has not yet
grown to a ``mature" size. Therefore, the resultant white dwarf will
appear with a small, ``premature" mass.  
In Figure~10, we plot the minimum white dwarf mass as a function of the orbital period and
eccentricity, log~[$P(1-e)^{\frac{3}{2}}$]. We also
show the model predictions of \markcite{rappaport95}Rappaport et al.\ (1995)
using their coefficient 
$R_{\rm 0}$ to be 3300~\rsun, consistent with the
Population~II case that they discussed.
If all the
secondary stars were white dwarfs and had evolved via stable mass
transfer, their minimum masses, plotted in Figure~10, should all lie to
the left of the theoretical curve, and we find that the
data do so, more or less, except for G202-65. Thus almost all of the
blue stragglers in Table~5, with orbital periods longer than 20 days, 
have orbital periods and eccentricities consistent with stable
mass transfer. This includes CS~$22956-028$, the star that shows
signs of $s$-process enhancements and an implied AGB donor star.
We show in Figure~10 the locations of the other two such stars
from SP2003, CS~$29497-030$ and CS~$29509-027$. All three stars
lie well within the regime of stable mass transfer.

\begin{figure}
\epsscale{1}
\figurenum{10}
\plotone{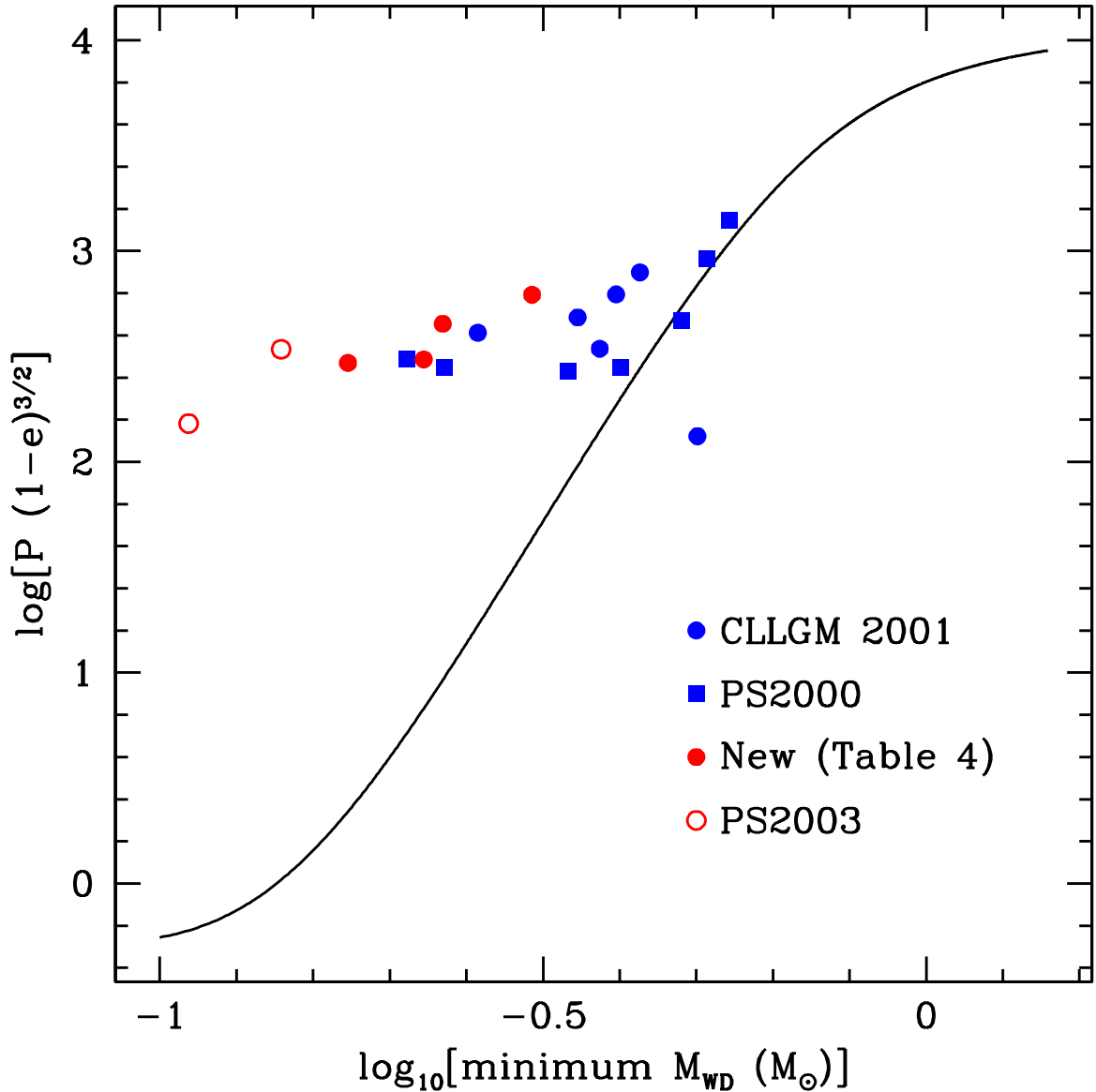}
\caption{The observed orbital periods
of blue straggler binary stars from this paper (red filled circles),
PS2000 (filled blue squares), and Carney et al.\ (2001;
blue filled circles) are compared with theoretical predictions for
stable mass transfer via Roche lobe overflow from
Rappaport et al.\ (1995). We have also included the two
other $s$-process enhanced stars CS~$29497-030$ and CS~$29509-27$
from SPC2003 as red open circles.
\label{fig10}}
\end{figure}

\placefigure{fig10}

\section{SUMMARY \& FUTURE WORK}

Our new search for radial velocity variability of five metal-poor
and lithium-deficient field stars has shown that four of them
are single-lined spectroscopic binaries with orbital periods
of a few hundred days and small orbital eccentricities.
Combining these results with those of \markcite{PS00}PS2000 
and \markcite{CLLGM}CLLGM, we
have found that field metal-poor dwarfs hotter than comparable
metallicity globular cluster turn-offs may represent stars
with two very different origins. 

Many of the binary stars studied by \markcite{PS00}PS2000 and most, if not
all, of the binary stars studied by \markcite{CLLGM}CLLGM and in this paper
are distinguished from normal binary stars by their unusually
low lithium abundances and unusually low binary orbital 
eccentricites. We have shown that the secondaries appear to
have a unusually narrow range of masses, and are consistent
with white dwarfs of 0.55 \msun. Despite the long periods of
these binary systems, we find that their \vrotsini\ values
are, on average, higher than those of single stars with
comparable temperatures, suggesting spin-up as part of
a mass transfer event. Using the models of
\markcite{rappaport95}Rappaport et al.\ (1995), we argue that in all cases but
one (G202-65), the orbital periods and eccentricities are
consistent with stable mass transfer. 
In some cases, \markcite{SPC2003}SPC2003
found enhanced values of $s$-process abundances, signifying
mass transfer from an evolved AGB star, but not all blue
stragglers show such enhancements. We have not discovered
any correlation between the presence or lack of such
enhancements and the binary system orbital properties.

Another interesting result is that among the constant velocity
stars, lithium abundances are normal, including one star, BD+72~94,
that appears to be in the ``lithium gap". HD~142575, on the
other hand, may lie in the metal-poor equivalent lithium gap, which
may occur at somewhat higher temperatures than is seen in the Hyades.
More lithium
abundance determinations for constant velocity stars in
this temperature regime would be very interesting, both
to strengthen the use of lithium as a distinguishing feature
of blue stragglers created by mass transfer, and as a tool
for exploring the cause of the lithium gap as well. 
These constant velocity stars
may have originated within a dwarf satellite galaxy that 
underwent star formation over a long period of time and which
only recently merged with the Milky Way, as suggested by \markcite{PBS94}PBS.

Additional radial velocities should be obtained of the constant
velocity and binary stars in our samples (\markcite{CLLGM}CLLGM and this paper)
as well as those stars studied by \markcite{PBS94}PBS, 
\markcite{PS00}PS2000, and \markcite{SPC2003}SPC2003
to refine the spectroscopic orbital solutions and more
securely identify the true constant velocity stars.

Finally, additional detailed abundance work should be undertaken
of these stars as well, following up on the fascinating
results of \markcite{PS00}PS2000 and, especially, \markcite{SPC2003}SPC2003.

It is a pleasure to thank the National Science Foundation for
financial support to the University of North Carolina and to Bowling
Green State University during the many years that we have been
puzzling over the blue straggler phenomenon. 
As always, we are indebted
to the many people who made observations for this project, especially
Joe Caruso, Perry Berlind, Bob Davis, Robert Stefanik, Jim Peters,
Mike Calkins, Ed Horine, Joe Zajack, Skip Schwartz,
Willie Torres, Ale Milone, and Dick McCroskey.

\clearpage

\begin{deluxetable}{lllllclcrcl}
\scriptsize
\tablenum{4}
\tablewidth{0pc}
\tablecaption{Stellar Data \label{tab:data}}
\tablenum{1}
\tablehead{
\colhead{Star}  &
\colhead{$B-V$} &
\colhead{[Fe/H]}   &
\colhead{T$_{\rm eff}$}    &
\colhead{P} &
\colhead{Gap?} &
\colhead{Li\tablenotemark{a}}    &
\colhead{Ref}    &
\colhead{[X/Fe]\tablenotemark{b}}    &
\colhead{Ref}    &
\colhead{Template}  }
\startdata
BD+23 74\tablenotemark{c} & 0.19 & $-0.91$ & 7500 & 837 & No & $< 1.32$ & 1 & \nodata &   & $8000/4.0/-1.0/30$ \nl
HD 8554\tablenotemark{d} & 0.29 & $-1.46$ & 6780 & 302 & Yes & $< 1.11$ & 1 & \nodata &   & $7250/4.0/-1.5/12$ \nl
HD 109443\tablenotemark{e} & 0.36 & $-0.55$ & 6650 & 724 & Yes & $< 0.67$ & 1 & \nodata & & $6500/4.0/-0.5/30$ \nl
HD 135449\tablenotemark{f} & 0.32 & $-0.92$ & 6740 & 327 & Yes & $< 1.1$ & 1,14 & \nodata & & $6250/4.0/-1.0/30$ \nl
HD 142575\tablenotemark{g} & 0.33 & $-0.97$ & 6550 & \nodata & Yes & 1.45 & 1,3 & \nodata;+0.05 & 3 & $6250/4.0/-1.5/14$\nl
         &  & & & & & & &  & & \nl
BD+72 94\tablenotemark{h} & 0.32 & $-1.62$ & 6620 & \nodata & Yes & 2.42 & 2 & \nodata & & $6250/4.0/-1.5/4$ \nl
BD+40 1166\tablenotemark{i} & 0.42 & $-0.76$ & 6200 & \nodata & No & \nodata &    & \nodata & & $6250/4.0/-1.0/8$ \nl
BD+25 1981 & 0.29 & $-1.26$ & 6860 & \nodata & No & $< 0.72$ & 1 & \nodata & & $6750/4.0/-1.5/10$ \nl
BD$-12$ 2669 & 0.28 & $-1.49$ & 6710 & 381 & Yes & \nodata &  & \nodata & & $7000/4.0/-1.5/30$\\
HD 84937 & 0.37 & $-2.18$ & 6300 & \nodata & No & 2.23 & 10,11 & +0.57/$-0.03$ & 3,8,10 & $6250/4.0/-2.0/4$ \nl
HD 97916 & 0.415 & $-1.31$ & 6250 & 663 & No & $< 1.2$ & 2 & \nodata;+0.05 & 3 & $6250/4.0/-1.5/12$\nl
HD 106516 & 0.45 & $-0.87$ & 6120 & 844 & No & $< 1.3$ & 7,12,13 & \nodata;$-0.83$ & 4 & $6000/4.0/-1.0/10$ \nl
BD+51 1817\tablenotemark{j} & 0.38 & $-1.10$ & 6330 & 517 & No & $< 1.64$ & 2 & \nodata & & $6250/4.0/-1.0/8$ \nl
G66-30 & 0.37 & $-1.75$ & 6280 & 688 & No & $< 1.61$ & 2 & $-0.09$;+0.16 & 15 & $6250/4.0/-1.5/6$ \nl
G202-65 & 0.36 & $-1.50$ & 6560 & 167 & Yes & $< 1.67$ & 2 & \nodata & & $6500/4.0/-1.5/10$ \nl
         &  & & & & & & & & & \nl
$22166-041$ & 0.16 & $-1.30$ & 6560 & 486 & Yes & \nodata & & $-0.21$;+0.46 & 5 & \nl
$22170-028$ & 0.17 & $-0.68$ & 8050 & 1.0 & No &  \nodata &   & \nodata &  & \nl
$22873-139$ & 0.34 & $-2.85$ & 6420 & 19 & Yes & 2.15 & 2 & \nodata &  & \nl
$22876-008$ & 0.28 & $-1.88$ & 6630 & 303 & Yes & \nodata & & +0.27;$-0.05$ & 5 & \nl
$22890-069$ & 0.20 & $-2.00$ & 7700 & 2.0 & No &  \nodata & & \nodata &   & \nl
$22892-027$ & 0.30 & $-1.03$ & 6720 & 485 & Yes & \nodata & & +0.26;+0.15 & 5 & \nl
$22948-068$ & 0.31 & $-1.37$ & 6590 & 300 & Yes & \nodata & & +0.15;+0.16 & 5 & \nl
$22956-028$ & 0.34 & $-2.08$ & 6900 & 1307 & No &  \nodata & & +1.38;+0.37 & 6 & \nl
$22966-054$ & 0.28 & $-1.17$ & 6785 & 306 & Yes & \nodata & & +0.15;+0.19 & 5 & \nl
$29518-039$ & 0.30 & $-2.49$ & 6510 & 1576 & Yes & \nodata & & \nodata &   & \nl
\tablevspace{4pt}
\tablenotetext{a}{Li abundances are given as log~n(Li), where log~n(H) = 12.00}
\tablenotetext{b}{Values given are [Sr/Fe] and [Ba/Fe], respectively.}
\tablenotetext{c}{BD+23~74 is cited as SAO~74088 by \markcite{GPS94}Glaspey et al.\ (1994).}
\tablenotetext{d}{HD 8554 is cited as SAO~109871 by \markcite{GPS94}Glaspey et al.\ (1994).}
\tablenotetext{e}{HD 109443 is cited as SAO~180920 by \markcite{GPS94}Glaspey et al.\ (1994).}
\tablenotetext{f}{HD 135449 is cited as SAO~206470 by \markcite{GPS94}Glaspey et al.\ (1994).}
\tablenotetext{g}{HD 142575 is cited as SAO~121258 by \markcite{GPS94}Glaspey et al.\ (1994).}
\tablenotetext{h}{BD+72 94 is often cited as G245-32.}
\tablenotetext{i}{BD+40 1166 is often cited as G96-20.}
\tablenotetext{j}{BD+51 1817 is often cited as G177-23.}
\tablerefs{(1) \markcite{GPS94}Glaspey et al.\ 1994; (2) \markcite{R01a}Ryan et al.\ 2001; 
(3) \markcite{JFB2000}Fulbright 2000; (4) \markcite{EAGLNT93}Edvardsson et al.\ 1993; 
(5) \markcite{PS00}PS2000; (6) \markcite{SPC2003}SPC2003; 
(7) \markcite{SPS94}Spite, Pasquini, \& Spite 1994;
(8) \markcite{MG2000}Mashonkina \& Gehren 2000;
(9) \markcite{MAGAIN}Magain 1989; (10) \markcite{RNB99}Ryan, Norris, \& Beers 1999; 
(11) \markcite{T94}Thorburn 1994;
(12) \markcite{SMFS93}Spite et al.\ 1993; (13) \markcite{LHE91}Lambert, Heath, \& Edvardsson 1991;
(14) \markcite{CPPMTA}Cutispoto et al.\ 2002; (15) \markcite{NRBD97}Norris et al.\ 1997}
\enddata
\end{deluxetable}

\clearpage

\begin{deluxetable}{llrr}
\footnotesize
\tablewidth{0pc}
\tablenum{2}
\tablecaption{Radial Velocities}
\tablehead{
\colhead{Tel} & \colhead{HJD} &
\colhead{$v_{\rm rad}$}  & \colhead{$\sigma$}}
 
\startdata

\cutinhead{BD +23 74~~~~00:32:43.3 $+$24:13:21}
W & 2450630.7792 &    33.98 & 2.89 \\
W & 2450655.7600 &    23.57 & 2.44 \\
W & 2450670.8359 &    31.28 & 1.85 \\
W & 2450691.8055 &    31.59 & 1.93 \\
W & 2450713.8439 &    28.01 & 1.61 \\
W & 2450728.7021 &    28.16 & 2.42 \\
W & 2450745.5618 &    37.96 & 2.48 \\
W & 2450771.5908 &    32.73 & 3.33 \\
W & 2450791.6284 &    26.96 & 3.94 \\
W & 2450811.5603 &    25.90 & 3.57 \\

\enddata
\end{deluxetable}

\clearpage

\begin{deluxetable}{llrrrrrrrrr}
\scriptsize
\tablenum{3}
\tablecaption{Mean Velocities and Errors}
\tablehead{
\colhead{Star}  &
\colhead{$\alpha$ (J2000) $\delta$} &
\colhead{$N_{\rm obs}$}   &
\colhead{Span}    &
\colhead{$v_{\rm rot}$ sin $i$} &
\colhead{$v_{\rm rad}$}    &
\colhead{$\sigma$}    &
\colhead{E}    &
\colhead{I}    &
\colhead{E/I} &
\colhead{P($\chi^2$)} }

\startdata
 &  &  &  &  &  &  &  &  &  &  \\
\multicolumn{11}{c}{New Program Stars}\\
 &  &  &  &  &  &  &  &  &  &  \\
BD+23 74  & 00:32:43.3 $+$24:13:21 & 55 & 2558 & 29.6 &  32.52 & 0.60 & 4.47 & 2.27 & 1.97 &  0.000000 \nl
HD~8554  & 01:24:42.3 $+$07:00:05 & 32 & 1544 & 11.0 &  11.13 & 0.64 & 3.60 & 1.05 & 3.44 &  0.000000 \\
HD~109443 & 12:34:46.7 $-$23:28:32 & 43 & 2342 & 29.1 &  43.36 & 0.48 & 3.14 & 1.25 & 2.50 &  0.000000 \\
HD~135449 & 15:16:10.3 $-$32:53:33 & 33 & 2575 & 29.5 & $-42.00$ & 0.69 & 3.94 & 1.37 & 2.88 &  0.000000 \\
HD~142575 & 15:55:02.8 $+05$:04:12 & 16 & 4758 & 13.0 & $-64.98$  & 0.17 & 0.54 & 0.68 & 0.80 & 0.784932 \\
 &  &  &  &  &  &  &  &  &  &  \\
\multicolumn{11}{c}{Binary Stars from \markcite{CLLGM}Carney et al.\ (2001)}\\
 &  &  &  &  &  &  &  &  &  &  \\
BD$-12$~2669 & 08:46:39.6 $-13$:21:25 & 135 & 2097 & 33.3 & 41.60 & 0.26 & 6.72 & 1.59 & 4.23 & 0.000000 \\
HD~97916 & 11:15:54.2 +02:05:12 & 32 & 5083 & 14.9 & 61.04 & 0.16 & 4.12 & 0.72 & 5.74 & 0.000000 \\
HD~106516 & 12:15:10.5 $-10$:18:44 & 39 & 5181 & 11.5 & 4.41 & 0.09 & 4.84 & 0.50 & 9.79 & 0.000000 \\
BD+51~1817 & 13:08:39.1 +51:03:59 & 28 & 2194 & 9.0 & $-58.64$ & 0.21 & 5.94 & 0.73 & 8.17 & 0.000000 \\
G66-30 & 14:50:07.8 +00:50:27 & 27 & 4366 & 8.1 & $-115.10$ & 0.14 & 4.50 & 0.81 & 5.55 & 0.000000 \\
G202-65 & 16:35:58.5 +45:51:59 & 49 & 3871 & 11.6 & $-245.60$ & 0.26 & 11.35 & 0.89 & 12.80 & 0.000000 \\
 &  &  &  &  &  &  &  &  &  &  \\
\multicolumn{11}{c}{Constant Velocity Stars from \markcite{CLLGM}Carney et al.\ (2001)}\\
 &  &  &  &  &  &  &  &  &  &  \\
BD+72~94 & 01:47:12.3 +73:28:27 & 40 & 7441 & 5.5 & $-268.67$ & 0.12 & 0.79 & 0.67 & 1.17 & 0.060699 \\
BD+40~1166 & 05:05:28.7 +40:15:26 & 38 & 7997 & 8.6 & 105.31 & 0.12 & 0.75 & 0.66 & 1.13 & 0.067409\\
BD+25~1981 & 08:44:24.6 +24:47:47 & 81 & 6909 & 10.0 & 57.54 & 0.08 & 0.75 & 0.57 & 1.33 & 0.000139\\
HD~84937 & 09:48:56.0 +13:44:39 & 82 & 7355 & 4.8 & $-15.38$ & 0.10 & 0.91 & 0.73 & 1.24 & 0.005681 \\
\enddata
\end{deluxetable}

\clearpage

\begin{deluxetable}{lrrrrrrrrrc}
\footnotesize
\tablenum{4}
\tablewidth{0pc}
\tablecaption{Orbital Parameters \label{orbparams}}
\tablehead{
\colhead{Star} &
\colhead{$P$} &
\colhead{$\gamma$} &
\colhead{$K$} &
\colhead{$e$} &
\colhead{$\omega$} &
\colhead{$T_{\rm 0}$} &
\colhead{$a_1 \sin i$} &
\colhead{$f(M)$} &
\colhead{$N$} &
\colhead{$\sigma$(o-c)} \\
\colhead{} &
\colhead{(days)} &
\colhead{(\kms)} &
\colhead{(\kms)} &
\colhead{} &
\colhead{(\arcdeg)} &
\colhead{} &
\colhead{Gm} &
\colhead{(\msun)} &
\colhead{Span} &
\colhead{(\kms)}}
\startdata

BD+23 74 &  840.4  &  +32.71 &   5.21 &   0.168  &    35  & 51974.6  &   59.4   &   0.0118  &   55 & 2.5  \\
              &\p 14.8  &\p  0.36 &\p 0.50 &\p 0.099 &\p  32  &\p  72.5  &\p 30.0   &\p 0.0171  & 2558 &      \\

HD 8554 &  302.5 &  +11.86 &   5.11 &   0.026 &   222  & 51111.1  &   21.2 &   0.0042  &   32 & 0.9 \\
              &\p  1.6 &\p  0.17 &\p 0.25 &\p 0.047 &\p 103  &\p 85.9  &\p  1.8 &\p 0.0011  & 1544 &      \\

HD 109443 &  684.5   &  +42.48 &   4.36 &   0.109  &   99 & 51743.3  &  40.8  &   0.0058  &   42 & 1.3  \\
              &\p 7.8   &\p  0.26 &\p 0.34 & \p 0.079  &\p  39  &\p  73.1  &\p  4.0   &\p 0.0035  & 2342 &      \\

HD 135449 &  326.1 &$-$42.50 &   5.49 &   0.053 &   283  & 51630.1  &   24.6 &   0.0056  &   33 & 1.3  \\
              &\p  1.9 &\p  0.30 &\p 0.34 &\p 0.062 &\p  70  &\p  61.9  &\p  4.0 &\p 0.0027  & 2575 &      \\

\enddata
\end{deluxetable}

\clearpage

\begin{deluxetable}{lrrcccccc}
\footnotesize
\tablewidth{0pc}
\tablenum{5}
\tablecaption{Mass and $\sin i$ Estimates}
\tablehead{
\colhead{Star} &
\colhead{log $P$} &
\colhead{$e$} &
\colhead{[Fe/H]} &
\colhead{($B-V$)$_{0}$} &
\colhead{$T_{\rm eff}$} &
\colhead{$M_1$} &
\colhead{$M_2$ (min)} &
\colhead{$\sin i$} }

\startdata
             &      &         &      &      &        &      &       &       \\
\multicolumn{9}{c}{Previous results from \markcite{CLLGM}Carney et al.\ (2001)} \\
             &      &         &      &      &        &      &       &       \\
BD~$-12$~2669 & 2.586 & 0.071 & $-1.49$ & 0.28 & 6710 & 0.91 & 0.38 & 0.742 \\
HD~97916 & 2.822 & 0.042 & $-1.00$ & 0.41 & 6235 & 0.93 & 0.39 & 0.771 \\
HD~106516 & 2.926 & 0.041 & $-0.87$ & 0.45 & 6100 & 0.91 & 0.42 & 0.818 \\
BD~+51~1817 & 2.713 & 0.043 & $-1.10$ & 0.38 & 6340 & 0.94 & 0.35 & 0.702 \\
G66-30 & 2.838 & 0.293 & $-1.75$ & 0.38 & 6310 & 0.85 & 0.26 & 0.551 \\
G202-65 & 2.224 & 0.145 & $-1.50$ & 0.36 & 6340 & 0.87 & 0.50 & 0.935 \\
CS~22166-041 & 2.687 & 0.024 & $-1.32$ & 0.32 & 6560 & 0.96 & 0.48 & 0.901 \\
CS~22876-008 & 2.481 & 0.075 & $-1.37$ & 0.28 & 6630 & 0.97 & 0.34 & 0.685 \\
CS~22892-027 & 2.686 & 0.259 & $-1.03$ & 0.30 & 6720 & 1.05 & 0.21 & 0.508 \\
CS~22948-068 & 2.477 & 0.044 & $-1.88$ & 0.31 & 6590 & 0.89 & 0.24 & 0.503 \\
CS~22956-028 & 3.116 & 0.207 & $-2.08$ & 0.34 & 6335 & 0.83 & 0.52 & 0.955 \\
CS~22966-054 & 2.486 & 0.059 & $-1.17 $ & 0.28 & 6785 & 1.04 & 0.40 & 0.776 \\
CS~29518-039 & 3.198 & 0.075 & $-2.49$ & 0.30 & 6510 & 0.86 & 0.55 & 1.000 \\
             &      &         &      &      &        &      &       &       \\
\multicolumn{9}{c}{New results} \\
             &      &         &      &      &        &      &       &       \\
BD+23~74 & 2.923 & 0.18 & $-0.91$ & 0.19 & 7500 & 1.28 & 0.31 & 0.619 \\
HD~8554 & 2.481 & 0.017 & $-1.46$ & 0.29 & 6780 & 0.98 & 0.18 & 0.390 \\
HD~109443 & 2.860 & 0.27 & $-0.55$ & 0.36 & 6650 & 1.18 & 0.22 & 0.471 \\
HD~135449 & 2.514 & 0.043 & $-0.92$ & 0.32 & 6740 & 1.08 & 0.21 & 0.447 \\

\enddata
\end{deluxetable}

\clearpage

\clearpage

\clearpage

\clearpage

\clearpage

\clearpage

\clearpage

\clearpage

\clearpage

\clearpage

\end{document}